\documentclass[sigconf]{acmart}

\usepackage[T1]{fontenc}
\usepackage{graphicx}
\usepackage{multirow}

\usepackage{subcaption}

\AtBeginDocument{%
  }

\copyrightyear{2023}
\acmYear{2023}
\setcopyright{rightsretained}
\acmConference[RAID '23]{The 26th International Symposium on Research in
Attacks, Intrusions and Defenses}{October 16--18, 2023}{Hong Kong, Hong
Kong}
\acmBooktitle{The 26th International Symposium on Research in Attacks,
Intrusions and Defenses (RAID '23), October 16--18, 2023, Hong Kong, Hong Kong}
\acmDOI{10.1145/3607199.3607242}
\acmISBN{979-8-4007-0765-0/23/10}

\usepackage{color}

\usepackage{pifont}
\newcommand{\cmark}{\text{\ding{52}}}

\newcommand{\reveal}{\textsc{ReVeal}}
\newcommand{\ours}{\textsc{DiverseVul}}

\begin{document}
\title{\textsc{DiverseVul}: A New Vulnerable Source Code Dataset for
Deep Learning Based Vulnerability Detection}

%
%

\author{Yizheng Chen}
\affiliation{%
  \institution{University of Maryland}
  \country{}
}
\email{yzchen@umd.edu}

\author{Zhoujie Ding}
\affiliation{%
  \institution{UC Berkeley}
  \country{}
}
\email{zhoujie.ding@berkeley.edu}

\author{Lamya Alowain}
\affiliation{%
  \institution{King Abdulaziz City for Science and Technology}
  \country{}
}
\email{lalowain@kacst.edu.sa}

\author{Xinyun Chen}
\affiliation{%
  \institution{Google Deepmind}
  \country{}
}
\email{xinyunchen@google.com}

\author{David Wagner}
\affiliation{%
  \institution{UC Berkeley}
  \country{}
}
\email{daw@cs.berkeley.edu}

\begin{abstract}

We propose and release a new vulnerable source code dataset. We curate the dataset by crawling
security issue websites, extracting vulnerability-fixing commits and source codes
from the corresponding projects.
Our new dataset contains 18,945 vulnerable functions spanning 150 CWEs and 330,492 non-vulnerable functions extracted from 7,514 commits.
Our dataset covers 295 more projects than all previous datasets combined.

Combining our new dataset with previous datasets,
we present an analysis of the challenges and promising research directions
of using deep learning for detecting software vulnerabilities.
We study 11 model architectures belonging to 4 families.
Our results show that deep learning is still not ready for vulnerability detection,
due to high false positive rate, low F1 score, and difficulty of detecting hard CWEs.
In particular, we demonstrate an important generalization challenge for
the deployment of deep learning-based models.
We show that increasing the volume of training data may not further improve the performance of deep learning models for vulnerability detection, but might be useful to improve the generalization ability to unseen projects.

We also identify hopeful future research directions. We demonstrate that
large language models (LLMs) are a promising research direction for ML-based vulnerability detection,
outperforming Graph Neural Networks (GNNs) with code-structure features in our experiments.
Moreover, developing source code specific pre-training objectives is a promising research
direction to improve the vulnerability detection performance.

\end{abstract}
%
%


\maketitle


\section{Introduction}

Detecting software vulnerabilities is crucial to prevent cybercrimes and economic losses,
but to date it remains a hard problem.
Traditional static and dynamic vulnerability detection techniques suffer from shortcomings.
Given the tremendous success of deep learning in image and natural language applications,
it is natural to wonder if deep learning can
enhance our ability to detect vulnerabilities~\cite{chakraborty2021deep,li2018vuldeepecker,zhou2019devign,russell2018automated,mirskyvulchecker}.
However, as we show in this paper, we still need to overcome many challenges before deep learning can achieve great performance
for vulnerable source code detection.

For deep learning to be successful, we need a large dataset of vulnerable source code. We release a new open vulnerability dataset for C/C++, \ours{}.
To curate the dataset, we crawl security issue websites, collect vulnerability reports, extract vulnerability-fixing commits for each vulnerability, clone the corresponding projects,
and extract vulnerable and nonvulnerable source code from them.
Our dataset contains 18,945 vulnerable functions and 330,492 nonvulnerable functions
extracted from 7,514 commits, covering 150 CWEs. This is more than twice the size of the C/C++ data from the previous largest and most diverse dataset CVEFixes~\cite{bhandari2021cvefixes}.
Our dataset is more diverse and covers almost 50\% more projects than the combination of all previously published datasets.
We publicly release the \ours{} dataset to the community at \url{https://github.com/wagner-group/diversevul}.

\begin{figure}[t!]
    \centering
    \includegraphics[width=0.48\textwidth]{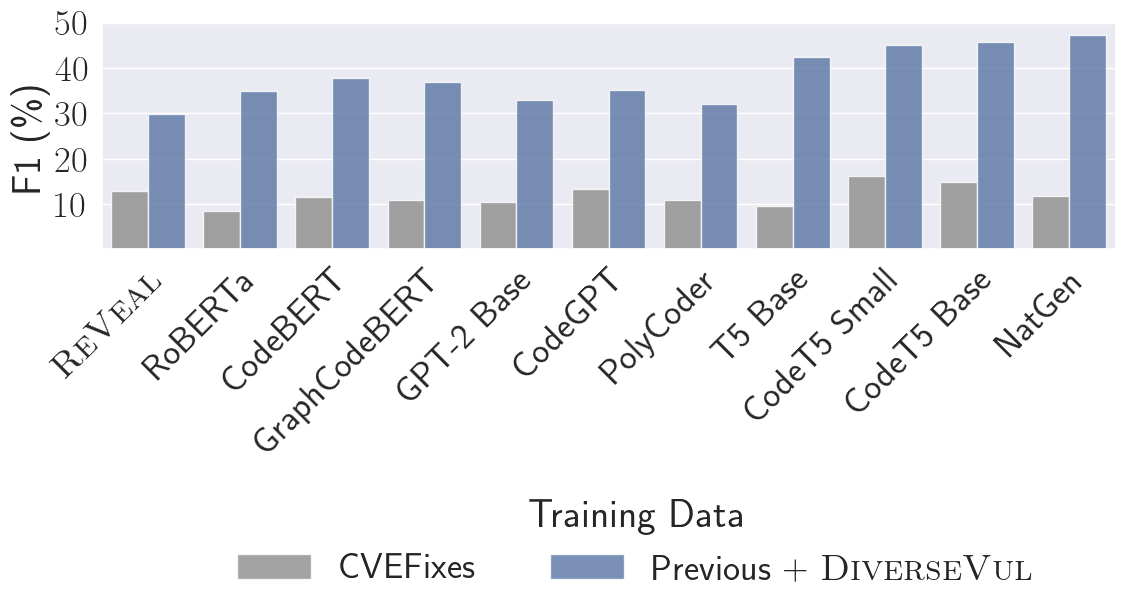}
    \caption{An overview of several of our results. When trained on only the CVEFixes dataset, \reveal{} has comparable performance as large language models.
    If we have enough data (Previous + \ours{}), large language models (e.g., NatGen) are superior to previous-generation models (e.g., \reveal{}, a GNN model with code-structure features), but we need large datasets to see these benefits.
    LLMs are better able to take advantage of larger datasets than previous-generation models (blue bars vs gray bars). 
    The best LLMs for this task, CodeT5 and NatGen, have been pre-trained with code-specific tasks.}
    \label{fig:f1_bars_ex_prev}
\end{figure}

Our new dataset has enabled us to study the state-of-the-art deep learning methods and gain new insights about promising research directions as well as the challenges for ML-based vulnerability detection.
In particular, we study several questions.
Does more training data help, or are models saturated?
Does the model architecture make a big difference?
Is it better to use the state-of-the-art model that relies on code-structure features, or better to use large language models?
Is a larger LLM better than a smaller LLM?
What are the most promising directions for further improving deep learning for vulnerability detection?

To study the effect of model architectures, we experiment with 11 different deep learning architectures from 4 representative model families: Graph Neural Networks (GNN)~\cite{li2015gated}, RoBERTa~\cite{liu2019roberta,feng2020codebert,guo2020graphcodebert}, GPT-2~\cite{radford2019language,lu2021codexglue,xu2022systematic}, and T5~\cite{raffel2020exploring,wang2021codet5,chakraborty2022natgen}. Much work on deep learning for vulnerability detection used GNNs with code-structure features~\cite{chakraborty2021deep,zhou2019devign,mirskyvulchecker}. We also explore applying large language models (LLMs) to vulnerability detection, as LLMs have achieved  state-of-the-art results for natural language processing and code understanding, even though they don't use code-structure features.
We study the performance of these models on three datasets:
(1) CVEFixes~\cite{bhandari2021cvefixes}, the largest previously published dataset of C/C++ vulnerabilities;
(2) the combination of all previously published datasets (Devign~\cite{zhou2019devign}, \reveal{}~\cite{chakraborty2021deep}, BigVul~\cite{fan2020ac}, CrossVul~\cite{nikitopoulos2021crossvul}, CVEFixes~\cite{bhandari2021cvefixes}), deduplicated;
(3) the combination of those  previous datasets and our \ours{} (details in Table~\ref{tab:mergedata}). 

Our experiments show that, when evaluating on a prior dataset CVEFixes~\cite{bhandari2021cvefixes}, the model architecture has little effect and LLMs perform about the same as GNNs.
In particular, on CVEFixes, the largest previously released dataset, the \reveal{} model (a GNN) achieves 12.8 F1 score, vs F1 scores of 8.5--16.3 for LLMs (see Figure~\ref{fig:f1_bars_ex_prev}).
One might be tempted to conclude from this that the exact architecture has little effect.
However, when evaluating on larger datasets, we can see that this conclusion is reversed: LLMs can perform significantly better than GNNs.
In particular, when we combine all previously published datasets together with our \ours{}, the best LLM achieves F1 score of 47.2, vs 29.8 for \reveal{}.

These experiments show that we need large datasets to reliably evaluate deep learning approaches to vulnerability detection, as the relative performance of different architectures shifts radically as we increase the amount of training data available:
a $5\times$ increase in the amount of training data (from CVEFixes to all datasets) improved the performance of our best model from 10.5 to 48.9 F1 score.
They suggest that LLMs are better able to make use of large datasets than GNNs: larger datasets improve the performance of \reveal{} only modestly, but improve the performance of LLMs significantly.
However, our experiments suggest that the performance gain from gathering more data may have stagnated. By adding our dataset to the combination of previous datasets, we can improve the test performance on 7 models out of 11. However, for the 3 best-performing models, either we don't see improvement or the improvement is small (details in Section~\ref{sec:add-our-data}).




Unfortunately, the state-of-the-art deep learning techniques are still not ready for vulnerability detection yet.
Our best model has 47.2\% F1 score, 43.3\% true positive rate, and 3.5\% false positive rate.
The false positive rate is still far too high for
the model to be practically useful.
A project might contain tens of thousands of functions, and this false positive rate corresponds to hundreds of false positives, which is more than most analysts are likely to be willing to wade through~\cite{bessey10}.


Despite the challenges, Figure~\ref{fig:f1_bars_ex_prev} suggests that large language models (LLMs) may be superior for deep learning based vulnerability detection.
In previous papers, researchers believe that GNN with code-structure features is promising for vulnerability detection~\cite{chakraborty2021deep,zhou2019devign,mirskyvulchecker},
since it combines domain knowledge with deep learning. In contrast, our results show that
large language models (RoBERTa, GPT-2, and T5 families) significantly outperform the state-of-the-art GNN, especially when training with more data.
In particular, CodeT5 models (CodeT5 Small, CodeT5 Base, NatGen) are the best.


Contrary to the common belief that model size is the most important factor for LLMs
to perform well, our results show that the most important factor may be how the LLM is trained.
Pretraining on code understanding tasks appears to offer large improvements.
For example, CodeT5 Small is pretrained to predict variable and function names, and it
can achieve an average of 8 percentage points higher F1 score than models that are twice its size but were not pretrained on code.
Surprisingly, we found that pretraining tasks that are effective for natural language do not help vulnerability detection much.
Instead, it appears we need code-specific pretraining tasks.
We think that developing better code-specific pretraining tasks is a promising
research direction for improving deep learning based vulnerability detection.

Moreover, we identify an important generalization challenge for the deployment
of deep learning based models.
To deploy a model we need to detect vulnerabilities from new software projects
that do not appear in the training set.
We found that deep learning models perform very poorly in this setting.
In particular, past work has split data into training and test sets by a random split of the vulnerabilities, without regard to which project each vulnerability appears in.
However, in practice, we often want to run a vulnerability detection tool on a new project, so there won't be any vulnerabilities from that project in the training set.
To evaluate the performance of deep learning in this setting, we  set aside a held-out set of projects, which we call ``unseen projects''; we train on vulnerabilities from the other projects (``seen projects''), and then test on vulnerabilities from unseen projects.
The performance of all models on unseen projects decreases significantly, e.g., from a F1 score of 49\% on seen projects to only 9.4\% on unseen projects.
The cause is unclear; perhaps the model is overfitting to patterns or coding idioms that are specific to the particular projects that appear in the training set.
This generalization failure is likely to be a significant barrier to deploying deep learning vulnerability detection in practice.
We hope future research will explore how to address this problem.
We suggest a simple intervention to use class weights in the training loss, that takes a small step in this direction, but the gap remains very large and more work is needed.

Lastly, we quantify the label noise in our dataset as well as previous datasets.
Label noise is a significant challenge for ML-based vulnerability detection research. To extract vulnerable functions from vulnerability-fixing commits, following the state-of-the-art approach (used by Devign~\cite{zhou2019devign}, \reveal{}~\cite{chakraborty2021deep}, BigVul~\cite{fan2020ac}, CrossVul~\cite{nikitopoulos2021crossvul}, CVEFixes~\cite{bhandari2021cvefixes}), we label functions that were changed by these commits as vulnerable. To understand the label accuracy of such labeling approach, we randomly sample 50 vulnerable functions from our dataset, and another 50 vulnerable functions from the union of three datasets that collect commits from NVD (BigVul, CrossVul, and CVEFixes). Then, we manually analyze the vulnerability and the labeled vulnerable functions. Our results find that the vulnerable function label in \ours{} is 60\% accurate, which is 24\% more accurate than the union of CVEFixes, BigVul and CrossVul but still containing many label errors. The main challenges are vulnerabilities that are spread across multiple functions and changes to non-vulnerable functions in vulnerability-fixing commits. We hope our work takes the first step towards understanding the label noise issue and highlights the need for deeper investigation of the impact of label noise.

We make the following contributions in this paper:
\begin{itemize}
\item We release \ours{}, a new C/C++ vulnerable source code dataset.
Our dataset is 60\% larger than the previous largest dataset for C/C++, and the most diverse
compared to all previous datasets.
\item We study 11 model architectures from 4 different model families. Our results show that
large language models outperform the state-of-the-art graph neural network for deep learning based vulnerability detection, and
developing source-code specific pretraining objectives is a promising research direction.
\item We identify challenges of deep learning for vulnerability detection. In particular,
we highlight the difficulty of generalizing to unseen projects outside the training set.
\item We assess label noise in our dataset and previous datasets that rely on vulnerability-fixing commits.
\end{itemize}


\section{Related Work}

In this section, we analyze previous public vulnerable source code datasets for C/C++, their labeling methods, and how they are used by related works on deep learning for vulnerability detection. 

\noindent\textbf{Synthetic Datasets:} 
SATE IV Juliet~\cite{okun2013report} and SARD~\cite{sard} are common synthetic datasets used by previous papers~\cite{russell2018automated,li2018vuldeepecker,mirskyvulchecker}. SARD expands on the Juliet v1.0 test suite and contains test cases for multiple programming languages. The test cases are highly accurate, and contain a variety of CWEs. However, they are constructed in isolation using known vulnerable patterns, which are designed to evaluate static and dynamic analysis tools. They don't fully capture the complexities of vulnerabilities within the real-world projects. The VulDeePecker~\cite{li2018vuldeepecker} dataset focuses on only two CWEs. They selected vulnerabilities from 19 projects according to CVE information from the National Vulnerability Database (NVD)~\cite{nvd}, and also combined SARD~\cite{sard} test cases from these two CWEs. Both VulDeePecker and SARD are semi-synthetic datasets.

\noindent\textbf{Static Analyzer Labels:}
The Draper~\cite{russell2018automated} dataset generated labels using alerts from three static analyzers: Clang, Cppcheck, and Flawfinder. Some categories of alerts were labeled as vulnerable, and some are mapped to non-vulnerable. The labeled dataset is at the function granularity. The quality of the label is unknown, but the label accuracy of static analyzers tend to be low.
D2A~\cite{zheng2021d2a} used differential analysis on the static analyzer (Infer) output over six open-source repositories. Given thousands of version pairs for a github repository, if the static analyzer generates an alert for the version before a git commit, but not after the commit, then D2A treats the commit as fixing a vulnerability. For the remaining alerts, D2A labels them as unrelated to vulnerabilities.

\noindent\textbf{Manual Labeling:}
The Devign~\cite{zhou2019devign} dataset was labeled by three security researchers. They first used keywords to find commits that likely fixed vulnerabilities and commits unrelated to vulnerabilities from four repositories. Then, for the first category, three security researchers manually reviewed these commits by majority vote to determine which fix security vulnerabilities. Given labels for each commit, Devign extracts the changed function before the commit as the data sample, and labels it as vulnerable or non-vulnerable according to the label of the commit. The authors of Devign released data for two repositories, \texttt{FFMPeg} and \texttt{Qemu}. This dataset has high quality labels, but manual labeling was very expensive, costing around 600 man-hours.

\noindent\textbf{Security Issues:}
Several prior datasets were generated by crawling security issues to identify vulnerability-fixing commits.
The \reveal{}~\cite{chakraborty2021deep} dataset was labeled using the patches to known security issues at Chromium security issues and Debian security tracker.
\reveal{} considers the changed functions before a security patch (commit) as vulnerable, after the patch as non-vulnerable, and all unchanged functions as non-vulnerable.
In comparison, our dataset \ours{} has 18K vulnerable functions, which is 11$\times$ the size of \reveal{} (Table~\ref{tab:mergedata}).

BigVul~\cite{fan2020ac}, CrossVul~\cite{nikitopoulos2021crossvul} and CVEfixes~\cite{bhandari2021cvefixes} collect vulnerability-fixing commits from Common Vulnerabilities and Exposures (CVE) records in the NVD~\cite{nvd}. In particular, CVEFixes covers all published CVEs up to 27 August 2022. CVEfixes and CrossVul datasets cover multiple programming languages, and we use their C/C++ data in this paper. These three datasets cover a wide range of projects and CWEs. In comparions, our dataset contains more projects, more CWEs, and double the number of vulnerability-fixing commits.

A few other vulnerable source code datasets in C/C++ do not provide vulnerable functions, and therefore we did not include them in our experiments. For example, AOSP~\cite{challande2022building} collected commits fixing CVEs from the security bulletin of Android Open Source Project (AOSP), which contain patches to vulnerabilities in Android OS, the Linux kernel, and system on chip manufacturers.
PatchDB~\cite{wang2021patchdb} provides patch information, i.e., code diffs, but does not provide enough information to identify the project or git repository it came from and thus does not let us reconstruct the full code of the changed funcction.

Security issues are effective at identifying vulnerability-fixing commits, as they are based on manual analysis from developers.
They are also representative of in-the-wild vulnerabilities in real-world projects. Therefore, we also collect our new dataset \ours{} by crawling security issues. Compared to all previous datasets, \ours{} is the most diverse one, covering the most number of projects. In particular, \ours{} has vulnerabilities from 295 new projects that have not been collected by any of the previous real-world datasets (Table~\ref{tab:mergedata}).

\noindent\textbf{DL for Vulnerable Source Code Detection:}
Previous papers have used LSTM~\cite{li2018vuldeepecker},
CNNs and RNNs~\cite{russell2018automated},
Bidirectional RNNS~\cite{li2021sysevr},
and Graph Neural Networks~\cite{chakraborty2021deep,zhou2019devign, mirskyvulchecker} to detect vulnerable source code.
A recent paper from Thapa et al.~\cite{thapa2022transformer} shows that
on the VulDeePecker~\cite{li2018vuldeepecker} dataset spanning two CWEs, large language models outperform BiLSTM and BiGRU models. However, they did not compare against Graph Neural Networks (GNN).
GNNs represent programs as graphs that contain useful domain knowledge for vulnerability detection. \reveal{}~\cite{chakraborty2021deep} used features obtained from the code property graph~\cite{yamaguchi2014modeling}, and VulChecker~\cite{mirskyvulchecker} proposed a new enriched program dependence graph. These papers used relatively small datasets such as \reveal{} and Juliet. If we train the models with larger datasets, it is not clear whether GNN with code-structure features is still effective compared to large language models.

\section{Data Collection}



Our goal is to collect high-quality vulnerability-fixing commits from a diverse set of real-world projects.
We focus on collecting data from security issues, since they reflect high-quality labels from a community of developers and security analysts. We start by identifying 29 security issue websites, and then narrow it down to 2 websites with most git system commits~\footnote{\texttt{snyk.io} and \texttt{bugzilla.redhat.com}.}. From these websites, we crawl the issue title, body, and relevant git commit URLs. Since developer's discussions may reference both vulnerability-fixing commits and vulnerability-introducing commits, we use two heuristics to exclude vulnerability-introducing commits. First, we exclude all commit URLs mentioned in comments containing keywords ``introduced" and ``first included"; and second, we manually go over all commits that changed at least 10 functions and exclude ones that introduced vulnerability. We keep the remaining commits in our dataset.

Next, we parse the git commit URLs to extract the projects and commit IDs.
We clone the projects and extract the commits from these projects.
We identify the C/C++ related code files in the commits. Then,
we extract all functions that were changed in these commits, and also functions
that did not change in the files. Same as \reveal{}~\cite{chakraborty2021deep}, 
we label the before-commit version of a changed function to be vulnerable, and the
after-commit version to be non-vulnerable. We label all unchanged functions in the related code files to be
non-vulnerable. Like prior work, we deduplicate functions by their MD5 hashes, and we do not normalize the code before deduplication. We keep track of the set of unique MD5s when processing the functions. We process all the vulnerable functions before the nonvulnerable ones. If the MD5 of a function already exists in this set, we do not include the function again in the data.
In total, we have collected 7,514 commits from 797 projects,
which result in 18,945 vulnerable functions and 330,492 non-vulnerable functions,
covering 150 CWEs. Table~\ref{tab:top_ten} shows the top 10 projects and the top 10 CWEs in \ours{} with the most number of vulnerability-fixing commits. Note that CWE-703 ``Improper Check or Handling of Exceptional Conditions" is not on the list of MITRE top-25 CWEs.

For issue titles that mention the CVE number, we query the National Vulnerability Database
API to obtain the CWE information for the issue and the corresponding commit.
For issues with developer annotated vulnerability category, we manually map them
to top 25 most popular CWEs. About 85\% of our data can be mapped to 150 CWE categories. We do not specifically address hierarchical CWEs. Depending on the query result from the NVD Database, a CVE number could be mapped to multiple CWEs.

\begin{table}[t]
    \begin{tabular}{l | r }
    \toprule
    {\bf Project} & {\bf \# Commits} \\
    \midrule
    linux & 1,458 \\
    ImageMagick & 330 \\
    php-src & 301 \\
    openssl & 261 \\
    tensorflow & 243 \\
    qemu & 205 \\
    linux-2.6 & 179 \\
    vim & 134 \\
    FFmpeg & 134 \\
    tcpdump & 112\\
    \bottomrule
    \multicolumn{2}{c}{\textbf{(a)}}
    \end{tabular}
    \hfill
    \begin{tabular}{l | r }
    \toprule
    {\bf CWE} & {\bf \# Commits} \\
    \midrule
    CWE-787 & 2,896 \\
    CWE-125 & 1,869 \\
    CWE-119 & 1,633 \\
    CWE-20 & 1,315 \\
    CWE-703 & 1,228 \\
    CWE-416 & 1,005 \\
    CWE-476 & 975 \\
    CWE-190 & 783 \\
    CWE-200 & 747 \\
    CWE-399 & 509 \\
    \bottomrule
    \multicolumn{2}{c}{\textbf{(b)}}
    \end{tabular}
\caption{Top 10 projects and CWEs in \ours{} and the corresponding number of vulnerability-fixing commits.}
\label{tab:top_ten}
\end{table}

\section{Experiments}

In this section, we study how our new dataset can improve
the performance of deep learning based vulnerability detection. We study 11 model architectures from 4 model families.
We also discuss insights learned from these experiments.

\subsection{Model Architectures}

We study 4 model families, where 3 families are transformer-based large language models (LLM). Within each LLM family, there are different variants of the model pretrained using different objectives.
Table~\ref{tab:modelsize} summarizes the number of parameters for all model architectures.

\subsubsection{Graph Neural Network}

Within the Graph Neural Network (GNN) family, we choose to
reproduce a representative previous work \reveal{}~\cite{chakraborty2021deep}.

Given a function, the \reveal{} model constructs a graph
to represent the function, computes the embedding vector of the graph, and classifies the vector as vulnerable or nonvulnerable. Specifically, the graph representation for the function is a code property graph~\cite{yamaguchi2014modeling} (CPG). CPG combines Abstract Syntax Tree (AST), Control Flow Graph (CFG), Data Flow Graph (DFG), and Program Dependence Graph (PDG). Each node has the corresponding source code and type, and each edge has a type. The embedding of the graph is a sum of embeddings of the nodes in the graph. To learn the node embeddings, \reveal{} uses Gated Graph Neural Networks (GGNN)~\cite{li2015gated} to recursively update the embeddings of the nodes. The initial embedding of a node is a concatenation of Word2Vec embedding of the code and the categorical type vector. Then, the GGNN training procedure uses the message passing mechanism to update each node embedding according to the node's neighbors in the graph. Finally, after training the GGNN, \reveal{} adds two fully-connected layers, rebalances the training set, to learn the final classifier. The total number of parameters of the \reveal{} model is 1.28M.

\begin{table}[!bt]
	\centering
	\begin{tabular}{c | c | r}
		\toprule
		\textbf{Model Family} & \textbf{Model Architecture} & \textbf{\# Parameters} \\ \midrule\midrule
		GNN & \reveal{} & 1.28M \\ \midrule \midrule
		\multirow{3}{*}{RoBERTa} 
	    & RoBERTa & 125M \\ \cmidrule{2-3}
	    & CodeBERT & 125M \\ \cmidrule{2-3}
		& GraphCodeBERT & 125M \\ \midrule \midrule
		\multirow{3}{*}{GPT-2} 
	    & GPT-2 Base & 117M \\ \cmidrule{2-3}
	    & CodeGPT & 124M \\ \cmidrule{2-3}
		& PolyCoder & 160M \\ \midrule \midrule
		\multirow{4}{*}{T5} 
	    & T5 Base & 220M \\ \cmidrule{2-3}
	    & CodeT5 Small & 60M \\ \cmidrule{2-3}
		& CodeT5 Base & 220M \\ \cmidrule{2-3}
		& NatGen & 220M \\ \bottomrule
	\end{tabular} 
	\caption{The number of parameters for different models.}
	\label{tab:modelsize}
\end{table}

\subsubsection{RoBERTa Family}

We select three model achitectures from the RoBERTa family: RoBERTa~\cite{liu2019roberta}, CodeBERT~\cite{feng2020codebert}, and GraphCodeBERT~\cite{guo2020graphcodebert}. All of them have 12 layers of Transformer encoders, 768 dimenional hidden states, 12 attention heads, and 125M model parameters in total.
The common pretraining objective for this family is masked language modeling (MLM). The MLM pretraining process randomly masks a percentage of tokens within the input tokens, effectively removing them, and the training goal is to predict the missing tokens.

RoBERTa~\cite{liu2019roberta} is an extension of BERT~\cite{devlin2018bert} that makes changes to important hyperparameters, including removing the pretraining objective of predicting the next sentence, as well as using larger mini-batches and learning rates during training. RoBERTa was pretrained on a union of five datasets: BookCorpus, English Wikipedia, CC-News, OpenWebText, and Stories.

CodeBERT~\cite{feng2020codebert} pretrains the model using the CodeSearchNet~\cite{husain2019codesearchnet} dataset containing 2.3M functions from six programming languages (Go, Java, JavaScript, PHP, Python, and Ruby). CodeBERT performs MLM pretraining and replaced token detection pretraining. During pretraining, each input is a pair of natural language description and source code, where the text describes the meaning of the code. The MLM pretraining in CodeBERT makes sure that tokens from both the natural language part and the source code part are masked out, and the replaced token detection corrupts both parts of the input as well. CodeBERT outperforms RoBERTa on two downstream tasks, natural language code search and code documentation generation.

GraphCodeBERT~\cite{guo2020graphcodebert} also uses the CodeSearchNet~\cite{husain2019codesearchnet} training datasets. In addition to having the natural language description and the source code parts of the input, GraphCodeBERT pretraining also constructs a third part of the input that captures the data flow between variables in the source code. In addition to MLM pretraining, GraphCodeBERT proposes two new pretraining objectives: edge prediction and node alignment. The edge prediction task maximizes the dot product between embeddings of two nodes if there is an edge, and the node alignment task maximize the dot product between embeddings of the code token and variable token if the variable represents the code token. Over benchmark datasets, GraphCodeBERT outperforms CodeBERT and RoBERTa on code clone detection, code translation, and code refinement tasks.

Note that the training dataset of CodeBERT and GraphCodeBERT does not have programs written in C/C++.

\subsubsection{GPT-2 Family}

We select three model architecures from the GPT-2 family: GPT-2 Base~\cite{radford2019language}, CodeGPT~\cite{lu2021codexglue}, and PolyCoder~\cite{xu2022systematic}. They have 12 layers of Transformer decoders, 768 dimentional hidden embeddings, and 12 attention heads. The size of the models are in Table~\ref{tab:modelsize}, ranging from 117M to 160M. The common pretraining objective for this family is causal language modeling, i.e., next token prediction. How well a model is pretrained on the causal language modeling is measured by perplexity. A lower perplexity value indicates a better model.

GPT-2~\cite{radford2019language} was pretrained on an unreleased WebText dataset, which was collected by scraping web page links on Reddit.

CodeGPT~\cite{lu2021codexglue} uses the same training objective and architecture of GPT-2, but different training data. The authors select Python and Java codes from CodeSearchNet~\cite{husain2019codesearchnet} as the training set, and release several variants of the pretrained CodeGPT models. In this paper, we use an adapted version of CodeGPT pretrained on Java codes. The CodeGPT model was initialized from GPT-2 weights, and then pretrained using Java codes from CodeSearchNet using the next token prediction task. Note that there is no C/C++ programs in the training set.

PolyCoder~\cite{xu2022systematic} uses the same model architecture and pretrianing objective as GPT-2, but pretrains the model from scratch. The authors pretrained the model with data from GitHub containing both source code and natural language comments within the code files. They cloned a total of 147,095 projects, that are the most popular repositories of 12 popular programming languages with at least 50 stars. Their training data contains over 24K repositories in C/C++. The authors curate an evaluation datasets of codes from unseen repositories. On C programming language, PolyCoder achieves the lowest perplexity value, compared to GPT-Neo, GPT-J, and Codex.

\subsubsection{T5 Family} We select four model achitectures from the T5 family: T5 Base~\cite{raffel2020exploring}, CodeT5 Base, CodeT5 Small~\cite{wang2021codet5}, and NatGen~\cite{chakraborty2022natgen}. All models have encoder-decoder Transformer layers. CodeT5 Small has 6 encoder layers and 6 decoder layers, 512 dimensional hidden states, 8 attention heads, and 60M parameters. The other models have 12 encoder layers and 12 decoder layers, 768 dimensional hidden states, 12 attention heads, and 220M parameters.

T5~\cite{raffel2020exploring} pretrains the model using the masked language modeling objective. In particular, T5 pretraining procedure randomly masks spans of tokens. The pretraining dataset is C4 (Colossal Clean Crawled Corpus). The authors curate the C4 dataset by processing the Common Crawl dataset to get hundreds of gigabytes of clean English text.

\begin{table*}[!bt]
	\centering
	\begin{tabular}{c | c | c | r | r | r | r }
		\toprule
		\textbf{Dataset} & \textbf{\# Projects} & \textbf{\# CWEs} & \textbf{\# Functions} & \textbf{\# Vul Func} & \textbf{\# Vul Func with CWE Info} & \textbf{\# Commits}\\ \midrule\midrule
            Devign & 2$^\triangledown$ & N/A & 26,037 & 11,888 & N/A & N/A \\ \midrule
            \reveal{} & 2$^\Diamond$ & N/A & 18,169 & 1,664 & N/A & N/A \\ \midrule
            BigVul & 348 & 91 & 264,919 & 11,823 & 8,783 & 3,754 \\ \midrule
            CrossVul$^*$ & 498 & 107 & 134,126 & 6,884 & 6,833 & 3,009 \\ \midrule
            CVEFixes$^*$ & 564 & 127 & 168,089 & 8,932 & 8,343 & 3,614 \\ \midrule
            \ours{} & \textbf{797} & \textbf{150} & \textbf{330,492} & \textbf{18,945} & \textbf{16,109} & \textbf{7,514} \\ 
            \midrule\midrule
		Previous\dag & 638 & 140 & 343,400 & 30,532 & 14,159 & 17,956 \\ \midrule
		Previous + \ours{} & \textbf{933} & \textbf{155} & \textbf{523,956} & \textbf{41,377} & \textbf{22,382} & \textbf{21,949} \\ \bottomrule
   \multicolumn{7}{l}{\dag: We aggregate previous five datasets by combining and deduplicating samples from Devign, \reveal{}, BigVul, CrossVul, and CVEfixes.}\\
   \multicolumn{7}{l}{$^*$: CVEfixes and CrossVul are multi-language datasets. We report numbers for C/C++ code.}\\
    \multicolumn{4}{l}{$^\triangledown$: Devign authors released data from two repositories: \texttt{FFMPeg}+\texttt{Qemu}.} & \multicolumn{2}{l}{$^\Diamond$: Chromium and Debian packages.} \\
        \end{tabular}
	\caption{Statistics about previous five datasets, \ours{}, merged Previous dataset, and Previous + \ours{}.}
\label{tab:mergedata}
\end{table*}

CodeT5~\cite{wang2021codet5} uses the same underlying transformer 
architecture as T5. We consider two model sizes in our experiments: CodeT5 Base and CodeT5 small. The CodeT5 Small is the smallest LLM, with one third the model size of other T5 based models, and roughly half the model size of RoBERTa and GPT-2 family models. CodeT5 was pretrained on on both CodeSearchNet data and additional C/C\# projects from GitHub. In addition to the masked span prediction objective, CodeT5 utilizes the knowledge about whether a token is an identifer (a variabel name or a function name) and designs two new pretraining tasks. The new pretraining tasks are, masked identifier prediction (masking all identifiers) and identifier tagging (predict whether a token is an identifier). 

NatGen~\cite{chakraborty2022natgen} proposes a new pretraining objective called ``naturalizing'' pretraining. The naturalizing pretraining is similar to a code editing process, that takes some weird synthetic code and tranform that into developer-readable code. The authors generate un-natural code by semantic preserving code transformations including adding dead code, changing a while loop to a for loop without variable initialization, renaming variables, and inserting confusing code element, etc. Then, the pretraining objective asks the model to naturlize the code to the original developer-friendly form. The NatGen model starts the pretraining from the CodeT5 Base weights, and then continues the pretraining process using their new pretraining objective. Doing well on the naturalizing pretraining objective requires the model to understand the code well. Compared to CodeT5, NatGen improves the performance over various downstream tasks such as code translation, text to code generation, and bug repair.

\begin{table*}[t]
\begin{tabular}{c | c | c | c | c | c | r  r  r  r  r }
\toprule
 \multirow{2}{*}{\begin{tabular}{@{}c@{}}{\bf Model}\\{\bf Family}\end{tabular}} &  \textbf{Model} & \textbf{Pretrain} &  \textbf{Pretrain} &  \textbf{Code-Specific} & \multirow{2}{*}{\bf Training Set} & \multicolumn{5}{|c}{\textbf{Test on Prev + \ours{}} (\%)} \\
& \textbf{Arch} & \textbf{on Code} & \textbf{on C/C++} & \textbf{Pretrain Task} &  & Acc & Prec & Recall &  F1 & FPR \\
\midrule
\midrule
\multirow{3}{*}{GNN} & \multirow{3}{*}{\reveal{}} &  &  &  & CVEFixes & 82.12 & 11.56 & 14.37 & 12.81 & 11.06  \\
 &  &  &  &  & Previous & 86.30 & 25.35 & 25.63 & 25.49 & 7.59 \\
 & & & & & Prev + \ours{} & 82.81 & 23.75 & 39.83 & 29.76 & 12.87 \\
\midrule
\midrule
\multirow{9}{*}{RoBERTa} & \multirow{3}{*}{RoBERTa} &  &  &  & CVEFixes & 91.71 & 34.24 & 4.85 & 8.50 & 0.80 \\
&  &  &  &  & Previous & 90.98 & 40.97 & 31.11 & 35.37 & 3.86 \\
 & & & & & Prev + \ours{} & 91.68 & 46.02 & 28.22 & 34.98 & 2.85 \\
\cmidrule{2-11}
  & \multirow{3}{*}{CodeBERT} & \multirow{3}{*}{\cmark} &  &  & CVEFixes & 91.62 & 35.64 & 6.98 & 11.67 & 1.09 \\
 &  &  &  &  & Previous & 91.07 & 41.83 & 32.20 & 36.39 & 3.86 \\
 & & & & & Prev + \ours{} & 90.48 & 39.25 & 36.54 & 37.85 & 4.87 \\
\cmidrule{2-11}
  & \multirow{3}{*}{GraphCodeBERT} & \multirow{3}{*}{\cmark} &  & \multirow{3}{*}{\cmark}  & CVEFixes & 91.76 & 38.28 & 6.35 & 10.89 & 0.88 \\
 &  &  &  &  & Previous & 91.65 & 45.71 & 27.61 & 34.43 & 2.83 \\
 & & & & & Prev + \ours{} & 90.32 & 38.18 & 35.51 & 36.79 & 4.96 \\
\midrule
\midrule
\multirow{9}{*}{GPT-2} & \multirow{3}{*}{GPT-2 Base} &  &  &  & CVEFixes & 91.45 & 31.02 & 6.37 & 10.57 & 1.22 \\
 & & & & & Previous & 91.80 & 46.62 & 23.46 & 31.21 & 2.32 \\
 & & & & & Prev + \ours{} & 91.73 & 46.18 & 25.71 & 33.03 & 2.58 \\
\cmidrule{2-11}
  & \multirow{3}{*}{CodeGPT} & \multirow{3}{*}{\cmark} &  &  & CVEFixes & 90.77 & 26.22 & 8.98 & 13.38 & 2.18 \\
   & & & & & Previous & 91.59 & 44.51 & 24.48 & 31.58 & 2.63 \\
 & & & & & Prev + \ours{} & 91.36 & 43.48 & 29.62 & 35.23 & 3.32 \\
\cmidrule{2-11}
  & \multirow{3}{*}{PolyCoder} & \multirow{3}{*}{\cmark} & \multirow{3}{*}{\cmark} &  & CVEFixes & 91.12 & 26.56 & 6.78 & 10.81 & 1.62 \\
 & & & & & Previous & 91.28 & 42.44 & 27.66 & 33.49 & 3.23 \\
 & & & & & Prev + \ours{} & 91.97 & 48.76 & 23.78 & 31.96 & 2.15 \\
\midrule
\midrule
\multirow{12}{*}{T5} & \multirow{3}{*}{T5 Base} &  &  &  & CVEFixes & 91.57 & 32.23 & 5.65 & 9.61 & 1.02 \\
 & & & & & Previous & 92.15 & 50.80 & 32.15 & 39.38 & 2.68 \\
 & & & & & Prev + \ours{} & 91.96 & 49.14 & 37.17 & 42.33 & 3.32 \\
\cmidrule{2-11}
  & \multirow{3}{*}{CodeT5 Small} & \multirow{3}{*}{\cmark} & \multirow{3}{*}{\cmark} & \multirow{3}{*}{\cmark} & CVEFixes & 90.89 & 30.03 & 11.18 & 16.29 & 2.24 \\
 & & & & & Previous & 91.98 & 49.34 & 42.53 & 45.68 & 3.76 \\
 & & & & & Prev + \ours{} & 91.85 & 48.41 & 42.22 & 45.10 & 3.88 \\
\cmidrule{2-11}
  & \multirow{3}{*}{CodeT5 Base} & \multirow{3}{*}{\cmark} & \multirow{3}{*}{\cmark} & \multirow{3}{*}{\cmark} & CVEFixes & 91.41 & 34.76 & 9.39 & 14.79 & 1.52 \\
 & & & & & Previous & 92.16 & 50.68 & 42.46 & 46.20 & 3.56 \\
 & & & & & Prev + \ours{} & 92.11 & 50.36 & 41.81 & 45.69 & 3.55 \\
\cmidrule{2-11}
  & \multirow{3}{*}{NatGen} & \multirow{3}{*}{\cmark} & \multirow{3}{*}{\cmark} & \multirow{3}{*}{\cmark} & CVEFixes & 91.64 & 36.17 & 7.07 & 11.83 & 1.08 \\
 & & & & & Previous & 92.30 & 51.81 & 42.92 & 46.94 & 3.44 \\
 & & & & & Prev + \ours{} & \textbf{92.30} & \textbf{51.81} & \textbf{43.25} & \textbf{47.15} & \textbf{3.47} \\
\bottomrule
\end{tabular}
\caption{
We evaluate the models on the same test set from Previous + \ours{}. There isn't a big difference between model performance across different architectures if we only train on the CVEFixes dataset. However, if we train on larger datasets, large language models significantly outperform the GNN-based \reveal{} model. Among them, CodeT5 Small, CodeT5 Base, and NatGen models have the highest F1 scores. We highlight the row with the highest F1 score in bold. Pretraining the model using code-specific pretraining task over C/C++ is very effective.}
\label{tab:main}
\end{table*}


\subsection{Model Performance with More Data}
\label{sec:add-our-data}

\subsubsection{Dataset Setup}

Deep learning models perform well when they are trained on a lot of data. Therefore, we combine non-synthetic datasets with high-quality vulnerability labels from real-world projects, including Devign, \reveal{}, BigVul, CrossVul, and CVEFixes. We then combine them with \ours{} and remove duplicate samples to create the Previous + \ours{} dataset, as shown in Table~\ref{tab:mergedata}.

Table~\ref{tab:mergedata} presents the statistics for each of the previous five datasets, as well as our dataset, \ours{}, and the merged datasets. Compared to all previous datasets, \ours{} includes a larger number of projects, more CWEs, more vulnerable functions, and more vulnerability-fixing commits. Specifically, \ours{} contains 18,945 vulnerable functions, of which 16,109 have CWE information, more than twice the number in any previous dataset. Having more data associated with CWE information will provide us with a more comprehensive understanding of model prediction results. The last two rows in Table~\ref{tab:mergedata} show the unique new data provided by \ours{} in the merged datasets after deduplicating samples. Comparing Previous and Previous + \ours{} datasets, we can see that \ours{} contains 295 new projects that do not exist in any of the previous datasets. Moreover, \ours{} provides 10,845 unique new vulnerable functions. 

For our experiments, we randomly select 80\% of the samples from the Previous + \ours{} dataset as the training set, 10\% as the validation set, and 10\% as the test set. We also construct the Previous training and validation sets that only contain the previous five datasets, and training and validation sets that only contain CVEFixes data. This allows us to train models with different amounts of data and evaluate how much adding more data helps in improving the model's performance to predict the same test set from Previous + \ours{}.

\subsubsection{Results}

For each model architecture in Table~\ref{tab:modelsize}, we train three models, using CVEFixes, Previous, and Previous + \ours{} training datasets. We train the \reveal{} models from scratch, and we fine tune the large language models (LLMs) for the vulnerability detection task from pretrained model weights. This gives us 33 models in total.
The detailed training setups in our experiments can be found in Appendix~\ref{appendix:training_details}.

Table~\ref{tab:main} shows the performance of the models over the same test set from Previous + \ours{}. The following summarizes the results.

\textbf{Result 1: When trained on all available data, large language models significantly outperform the state-of-the-art GNN-based \reveal{} model.}
When trained on all available data (Previous + \ours{}), LLMs perform significantly better than the \reveal{} model: the \reveal{} model achieves a 29.76 F1 score, while LLMs achieve F1 scores from 31.96 to 47.15.
The best LLM performs significantly better than \reveal{} on this large training set.
Comparing between \reveal{} and LLMs is arguably unfair since ReVeal has 1--2 orders of magnitude fewer parameters than LLMs. We do not know whether a larger GNN could be competitive with LLMs.
Unfortunately,
even the best-performing model, NatGen, is not yet suitable for deployment in vulnerability detection, with a 3.47\% false positive rate and a 47.15\% F1 score.
This false positive rate is still too high to be practical, and the F1 score is still low. Nevertheless, we believe that large language models hold promise for deep learning-based vulnerability detection.

Interestingly, LLMs require a large amount of training data to surpass \reveal{}.
When trained solely on CVEFixes data, a much smaller training set, there is no clear advantage of LLMs over GNN-based ReVeal model, and ReVeal is even better than 6 LLMs (out of 10) in this setting.

\begin{figure}[t!]
    \centering
    \includegraphics[width=0.48\textwidth]{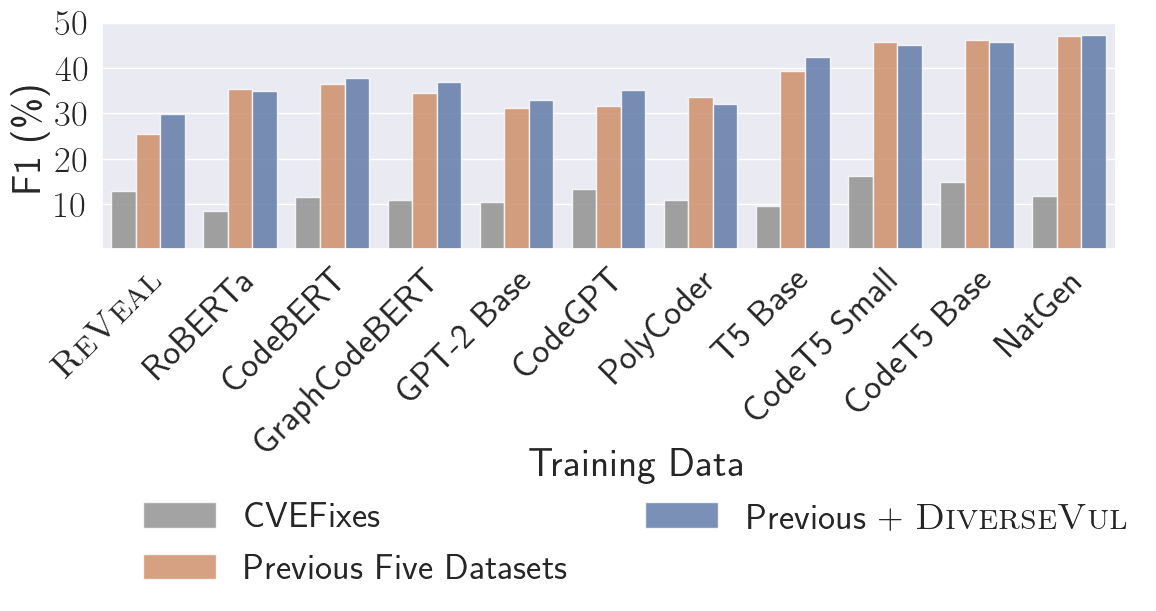}
    \caption{We visualize the performance of models that are trained on CVEFixes, Previous, and Previous + \ours{}. Adding \ours{} to the merged Previous dataset helps improve the test performance for 7 models out of 11. It does not help the CodeT5 models.}
    \label{fig:f1_bars}
\end{figure}

\textbf{Result 2: Within the three base LLM models, T5 Base performs better than RoBERTa and GPT-2 Base for vulnerability detection.} RoBERTa only uses encoders, GPT-2 only uses decoders, and T5 uses encoder-decoder Transformer layers. When trained on Previous + \ours{}, T5 Base has a test F1 score that is 7.35\% and 9.3\% higher than RoBERTa and GPT-2 Base, respectively. Thus, an encoder-decoder architecture might have an advantage over a decoder/encoder only architecture.

\textbf{Result 3: Pretraining on code does not lead to significant improvements in vulnerability prediction, if we only use natural language pretraining tasks.} The code models CodeBERT, GraphCodeBERT, CodeGPT, PolyCoder are not significantly better than the corresponding text models RoBERTa and GPT-2 Base. Specifically, when trained on the Previous dataset, CodeBERT and GraphCodeBERT perform similarly to RoBERTa. When trained on the Previous + \ours{} dataset, CodeBERT and GraphCodeBERT improve the F1 score by up to 2.8\% compared to RoBERTa. On the other hand, when trained on Previous dataset, CodeGPT and PolyCoder have up to 2.3\% higher F1 scores than GPT-2; but when trained on Previous + \ours{}, PolyCoder performs worse than GPT-2. Our findings suggest that pretraining models on code using MLM or next token prediction techniques does not yield significant improvements in detecting C/C++ vulnerabilities. While CodeBERT, GraphCodeBERT, and CodeGPT have not pretrained on C/C++, PolyCoder has pretrained over C/C++ code for next token prediction, which still does not help detecting C/C++ vulnerabilities.


\textbf{Result 4: Code-specific pretraining tasks on C/C++ make a big difference in improving vulnerability detection performance.} The two CodeT5 models and the NatGen model have the best F1 scores. They are pretrained using code-specific pretraining tasks on C/C++. CodeT5 models use identifier-aware pretraining tasks: masked identifier prediction and identifier tagging. NatGen does additional code naturalizing pretraining on top of CodeT5, such as removing dead code and renaming variables. These pretraining tasks ask the model learn about basic code understanding, which significantly improves the fine-tuned model performance for vulnerability detection task. Note that GraphCodeBERT also does some code-specific pretraining to learn embeddings from a pair of variables with data flow to have large dot product value. However, since it did not train on C/C++ data, it is unknown whether such pretraining task is effective for vulnerability prediction.

\textbf{Result 5: Code-specific pretraining task is more important than the model size.} Among the best three models in Table~\ref{tab:main} (CodeT5 Small, CodeT5 Base, NatGen), the CodeT5 Small model has only 60M parameters, half of the size of RoBERTa models and GPT-2 models, and less than one third the size of other T5 models. However, CodeT5 Small performs very similar to the largest CodeT5 Base and NatGen models, and it performs better than all the other models. Contrary to the belief that larger models tend to produce better performance, our results show that code-specific pretraining task is more important than the model size for vulnerability detection.

\begin{figure}[t!]
    \centering
    \includegraphics[width=0.4\textwidth]{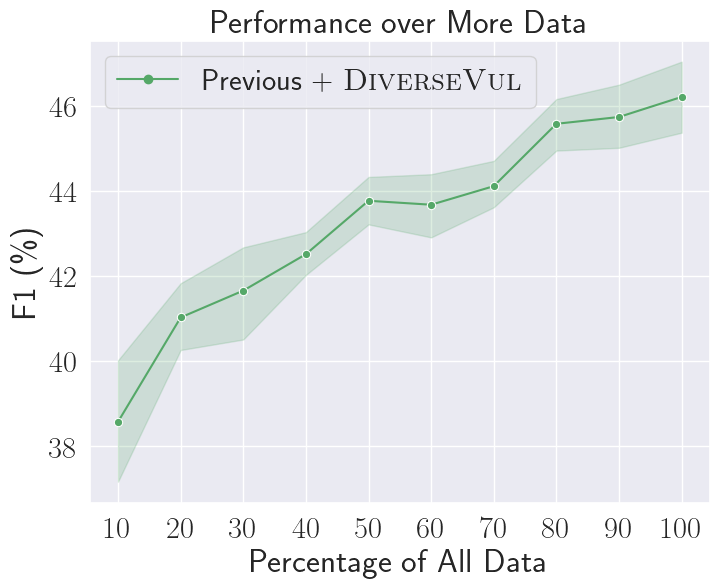}
    \caption{Deep learning for vulnerable source code detection benefits from more data collected from the same distribution as the test data.
    We fine-tune CodeT5 Small models on different amounts of vulnerable source code data with different volume and report the test F1 score.
    We run each dataset setup 10 times. The lines are the average,
    and the region denotes 95\% confidence interval.
    This figure shows that a larger training set improves the F1 score on vulnerability detection on test data from the same distribution.}
    \label{fig:volume}
\end{figure}

\textbf{Result 6: Performance gain from collecting more datasets may have saturated.}
Figure~\ref{fig:f1_bars} visualizes how much 
training on \ours{} + Previous data helps improve the vulnerability detection performance, compared to Previous data. Adding \ours{} to the training set improves the F1 score for 7 models by 2.4\% on average, compared to only training with the Previous dataset. However, it does not help the best performing CodeT5 models, and it only helps NatGen modestly. Even though we see a big improvement to model performance by training on the merged Previous datasets compared to only training on CVEFixes, collecting a different dataset may not further improve that.

\subsection{Dataset Volume}
\label{sec:volume}

\subsubsection{Dataset Setup}


\begin{table*}[t]
\begin{tabular}{c | c | c | c | c | c | r  r  r  r  r }
\toprule
 \multirow{2}{*}{\begin{tabular}{@{}c@{}}{\bf Model}\\{\bf Family}\end{tabular}} &  \textbf{Model} & \textbf{Pretrain} &  \textbf{Pretrain} &  \textbf{Code-specific} & \multirow{2}{*}{\bf Training Set} & \multicolumn{5}{|c}{\textbf{Test on Unseen Projects} (\%)} \\
& \textbf{Arch} & \textbf{on Code} & \textbf{on C/C++} & \textbf{Pretrain task} &  & Acc & Prec & Recall &  F1 & FPR \\
\midrule
\midrule
\multirow{2}{*}{GNN} & \multirow{2}{*}{\reveal{}} &  &  &  & Previous & 82.88 & 5.06 & 20.92 & 8.15 & 14.78 \\
 & & & & & Prev + \ours{} & 85.88 & 5.67 & 18.46 & 8.67 & 11.58 \\
\midrule
\midrule
\multirow{6}{*}{RoBERTa} & \multirow{2}{*}{RoBERTa} &  &  &  & Previous & 94.69 & 6.20 & 3.23 & 4.25 & 1.85 \\
 & & & & & Prev + \ours{} & 95.59 & 10.46 & 2.78 & 4.40 & 0.90 \\
\cmidrule{2-11}
  & \multirow{2}{*}{CodeBERT} & \multirow{2}{*}{\cmark} &  &  & Previous & 94.94 & 9.53 & 4.57 & 6.17 & 1.64 \\
 & & & & & Prev + \ours{} & \textbf{94.19} & \textbf{13.34} & \textbf{10.80} & \textbf{11.94} & \textbf{2.65} \\
\cmidrule{2-11}
  & \multirow{2}{*}{GraphCodeBERT} & \multirow{2}{*}{\cmark} &  & \multirow{2}{*}{\cmark} & Previous & 95.32 & 4.64 & 1.45 & 2.21 & 1.12 \\
 & & & & & Prev + \ours{} & 94.74 & 12.48 & 7.35 & 9.25 & 1.95 \\
\midrule
\midrule
\multirow{6}{*}{GPT-2} & \multirow{2}{*}{GPT-2 Base} &  &  &  & Previous & 94.92 & 6.19 & 2.78 & 3.84 & 1.60 \\
 & & & & & Prev + \ours{} & 95.06 & 9.82 & 4.34 & 6.02 & 1.51 \\
\cmidrule{2-11}
  & \multirow{2}{*}{CodeGPT} & \multirow{2}{*}{\cmark} &  &  & Previous & 94.32 & 5.98 & 3.79 & 4.64 & 2.25 \\
 & & & & & Prev + \ours{} & 94.47 & 9.86 & 6.35 & 7.72 & 2.19 \\
\cmidrule{2-11}
  & \multirow{2}{*}{PolyCoder} & \multirow{2}{*}{\cmark} & \multirow{2}{*}{\cmark} &  & Previous & 95.41 & 8.54 & 2.67 & 4.07 & 1.08 \\
 & & & & & Prev + \ours{} & 92.73 & 10.25 & 12.81 & 11.39 & 4.24 \\
\midrule
\midrule
\multirow{8}{*}{T5} & \multirow{2}{*}{T5 Base} &  &  &  & Previous & 95.67 & 20.21 & 6.35 & 9.66 & 0.95 \\
 & & & & & Prev + \ours{} & 96.16 & 34.00 & 5.68 & 9.73 & 0.42 \\
\cmidrule{2-11}
  & \multirow{2}{*}{CodeT5 Small} & \multirow{2}{*}{\cmark} & \multirow{2}{*}{\cmark} & \multirow{2}{*}{\cmark} & Previous & 95.02 & 12.21 & 5.90 & 7.96 & 1.60 \\
 & & & & & Prev + \ours{} & 94.91 & 13.35 & 7.24 & 9.39 & 1.78 \\
\cmidrule{2-11}
  & \multirow{2}{*}{CodeT5 Base} & \multirow{2}{*}{\cmark} & \multirow{2}{*}{\cmark} & \multirow{2}{*}{\cmark} & Previous & 96.21 & 32.32 & 3.56 & 6.42 & 0.28 \\
 & & & & & Prev + \ours{} & 95.56 & 18.03 & 6.12 & 9.14 & 1.05 \\
\cmidrule{2-11}
  & \multirow{2}{*}{NatGen} & \multirow{2}{*}{\cmark} & \multirow{2}{*}{\cmark} & \multirow{2}{*}{\cmark} & Previous & 95.48 & 17.86 & 6.68 & 9.72 & 1.16 \\
 & & & & & Prev + \ours{} & 95.49 & 17.38 & 6.35 & 9.30 & 1.14 \\
\bottomrule
\end{tabular}
\caption{We randomly choose 95 projects as unseen projects for testing. The remaining projects are used for training. We train each model on seen projects and test them on unseen projects. We highlight the row with the highest F1 score in bold. Overall, the F1 scores show that these models have poor generalization performance on unseen projects. Adding \ours{} to Previous training set helps improve the generalization performance for all models except NatGen.
}
\label{tab:test_unknown}
\end{table*}

We want to measure the effect of data volume on model performance for vulnerability detection.
We run the following experiment ten times. For each run, we randomly split the Previous + \ours{} into training, validation, and test sets. Then, we simulate the effect of different data volume by subsampling the training and validation sets. Specifically, we randomly sample 10\% to 90\% of the training and validation data from the full training and validation data of Previous + \ours{}. Then, we train the models, and evaluate them on the same original test set without subsampling.

\subsubsection{Results}

We fine tune 100 CodeT5 Small models on different dataset setups from 10 experiment runs. Within each run, we evaluate the models on the same final test set from the Previous + \ours{}, and train 10 models by using different percentages of training and validation data. Figure~\ref{fig:volume} plots the average and 95\% confidence interval for the test F1 score, when a model is fine tuned from a corresponding dataset setup.

\textbf{Result 7: Increasing the volume of the training dataset from the same distribution helps vulnerability detection.}
Our results show that training on a larger dataset \emph{from the same distribution} can improve the test performance. Figure~\ref{fig:volume} shows an upward trend of better test F1 score as the volume of training data increases. If we know the test data distribution ahead of the model deployment time, collecting more training data from that distribution might further improve the performance on vulnerability detection.

\subsection{Generalization}
\label{sec:generalization}


\subsubsection{Dataset Setup}

In a real-world deployment scenario, a vulnerability detection model needs to predict vulnerable source code in new developer projects that it has not been trained on. Therefore, we would like to test a model's performance on unseen projects.

We randomly select 95 unique projects from the merged Previous dataset as the unseen projects test set, to evaluate all models in this experiment. Then, the remaining projects are treated as seen projects in both training set and validation set. For both Previous and Previous + \ours{} datsets, we randomly sample 90\% of the seen projects as the training set, and 10\% remaining projects are the validation set. The training and validation sets of Previous + \ours{} are supersets of these from Previous. 



\begin{table*}[t!]
\begin{tabular}{c | c | r  r  r  r  r  | r  r  r  r  r }
\toprule
  \multirow{3}{*}{\textbf{Model Arch}} & \multirow{3}{*}{\textbf{Scheme}} & \multicolumn{5}{|c}{\textbf{Train on Seen Projects}} & \multicolumn{5}{|c}{\bf Train on Random Samples} \\
   & & \multicolumn{5}{|c}{\textbf{Test on Unseen Projects} (\%)} & \multicolumn{5}{|c}{\textbf{Test on Random Samples} (\%)} \\
 & & Acc & Prec & Recall &  F1 & FPR & Acc & Prec & Recall &  F1 & FPR \\
\midrule
\midrule
\multirow{4}{*}{CodeBERT} & No weight & 94.19 & 13.34 & 10.8 & 11.94 & 2.65 & 90.48 & 39.25 & 36.54 & 37.85 & 4.87 \\
 & Project Balanced & 95.09 & 11.6 & 5.23 & 7.21 & 1.51 & 90.7 & 34.43 & 18.97 & 24.46 & 3.11 \\
 & Weighted Soft F1 Loss & 91.38 & 11.41 & 20.16 & 14.57 & 5.92 & 90.72 & 34.55 & 18.9 & 24.43 & 3.08 \\
 & Class Weight & 92.16 & 12.21 & 18.6 & 14.74 & 5.06 & 89.39 & 36.97 & 47.89 & 41.72 & 7.04 \\
\midrule
\multirow{4}{*}{PolyCoder} & No weight & 92.73 & 10.25 & 12.81 & 11.39 & 4.24 & 91.97 & 48.76 & 23.78 & 31.96 & 2.15 \\
 & Project Balanced & 94.37 & 8.33 & 5.46 & 6.59 & 2.27 & 90.77 & 30.08 & 12.33 & 17.49 & 2.47 \\
 & Weighted Soft F1 Loss & 93.17 & 11.24 & 12.69 & 11.92 & 3.79 & 89.88 & 36.07 & 35.72 & 35.9 & 5.46 \\
 & Class Weight & 89.76 & 9.84 & 22.16 & 13.63 & 7.68 & 86.48 & 29.19 & 49.36 & 36.68 & 10.32 \\
\midrule
 \multirow{4}{*}{CodeT5 Small} & No Weighting & 94.91 & 13.35 & 7.24 & 9.39 & 1.78 & 91.85 & 48.41 & 42.22 & 45.10 & 3.88 \\
 & Project Balanced Batch Sampler & 95.3 & 14.52 & 5.90 & 8.39 & 1.31 & 90.69 & 39.36 & 31.96 & 35.27 & 4.24 \\
 & Weighted Soft F1 Loss & 96.34 & 48.18 & 5.90 & 10.52 & 0.24 & 91.31 & 44.69 & 39.78 & 42.09 & 4.24 \\
 & Class Weights for Cross Entropy Loss & \textbf{93.87} & \textbf{16.95} & \textbf{17.48} & \textbf{17.21} & \textbf{3.24} & \textbf{89.57} & \textbf{39.80} & \textbf{61.33} & \textbf{48.28} & \textbf{7.99} \\
\bottomrule
\end{tabular}
\caption{Using class weights for cross entropy loss improves the generalization performance of models, when they are trained on seen projects and tested on unseen projects. Using class weights improves the unseen project test F1 score of CodeBERT from 11.94\% to 14.74\%, PolyCoder from 11.39\% to 13.63\%, and CodeT5 Small from 9.39\% to 17.21\%. Moreover, if the training and testing samples are drawn from the same distribution, using class weights also improves the test F1 score. We highlight the row with the highest F1 score in bold.}
\label{tab:weighting}
\end{table*}

\subsubsection{Results}

We train \reveal{} and fine tune each LLM on the seen projects training set from Previous and Previous + \ours{}, resulting in 22 models in total. We make sure that these models have been trained well, since they have achieved validation performance similar to training performance.
Table~\ref{tab:test_unknown} shows the test performance of these models over unseen projects.

The F1 scores of all models on unseen projects are very low. The best models are CodeBERT, PolyCoder, CodeT5 Small, CodeT5 Base models trained on Previous + \ours{}, and NatGen model trained on Previous seen projects. Adding \ours{} to Previous training set helps improve the generalization performance for all models except NatGen. One recent, concurrent work~\cite{steenhoek2022empirical} also observed a significant performance drop when testing on unseen projects. In our experiment, we have included hundreds of more projects in the training set than~\cite{steenhoek2022empirical}, but we still observe the poor generalization results.

\textbf{Result 8: There is a significant challenge for deep learning models to generalize to unknown test projects on the vulnerability detection task.}
A popular use case of AI for Code is the GitHub CoPilot, where the AI model suggests ways to complete code to developers when they are writing code. If AI for deep learning detection is also a coding assistant, it needs to suggest potential vulnerable functions a developer is writing, in a new project it has not been trained on. Alternatively, static analyzers can be used to examine vulnerabilities in different projects. In a similar use case, deep learning based detection model needs to analyze a new project (after development) it has not seen before. Both of these use cases require the deep learning model to have strong generalization performance to new projects, and it is an open research problem for the community to tackle.

\subsection{Weighting}

In this section, we investigate whether three simple weighting schemes can potentially improve the model's generalization performance to unseen test projects. The weighting schemes are the following.

\subsubsection{Project Balanced Batch Sampler}
Our idea is to make the model perform equally well on different projects. Therefore, we propose a batch sampler that is equally likely to sample from any project in the training set. If a project is picked, it then randomly sample from all functions belonging to the project.

\subsubsection{Weighted Soft F1 Loss}
Since we care about F1 score as the final performance metric, we would like explore if a different loss function helps with improving the generalization performance. We use normalized prediction probabilities (between 0 and 1) from the training samples to calculate true positives, true negatives, false positives, and false negatives, as in floating point numbers. Then, we use these to compute two F1 scores of predicting the positive label (vulnerable function) and the negative label (nonvulnerable functions) separately. The loss for the positive label is 1 - positive F1 score, and the loss for the negative label is 1 - negative F1 score. Finally, we give a higher weight to the first loss value, proportional to the ratio of nonvulnerable to vulnerable functions in the data. Then, we choose the corresponding loss value according to the ground truth class label as the final training loss.

\subsubsection{Class Weights for Cross Entropy Loss}
In this scheme, we still use cross entropy loss for training. We upweight the loss value for the positive class (vulnerable class), proportional to the ratio of nonvulnerable samples over vulnerable samples. We use the same loss value for the negative class.

\subsubsection{Results}

We follow the same project split dataset setup described in Section~\ref{sec:generalization}. We fine tune CodeBERT, PolyCoder, and CodeT5 Small models over the seen projects training set from Previous + \ours{} dataset, and test them on 95 unseen projects. For each model architecture, we use four schemes to fine tune four models: no weighting, project balanced batch sampler, weighted soft F1 loss, and class weights for cross entropy loss.
In addition, we fine tune another four models for each architecture using these schemes over a different data split, the random data split described in Section~\ref{sec:add-our-data}.

\begin{figure}[t!]
    \centering
    \includegraphics[width=0.46\textwidth]{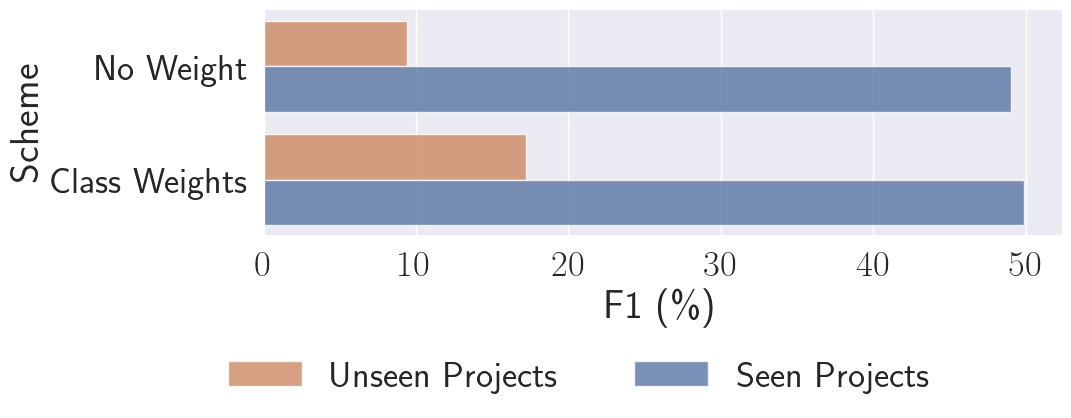}
    \caption{Using class weights in the training loss function improves the generalization performance over unseen projects for CodeT5 Small, and it slightly improves the performance on seen projects as well. The test F1 score on unseen projects is still quite low.}
    \label{fig:class_weights}
\end{figure}

\textbf{Result 9: Using class weights for cross entropy loss can improve the model's generalization performance to unseen projects, but there is a lot of room for further improvements. Class weights also improve the model's performance if training / testing samples are drawn from the same distribution.}
Table~\ref{tab:weighting} shows the evaluation results of models fine tuned with different schemes. For the seen / unseen projects experiment,
using class weights increases the F1 score for all three model architectures. The project balanced batch sampler does not help with generalization.
The weighted soft F1 loss helps CodeBERT and CodeT5 Small with generalization, but it hurts performance on seen projects.
Overall, class weights is the best scheme, as it improves performance on both seen and unseen projects. CodeT5 Small trained with class weights has the best test F1 score (17.21\%) on unseen projects.


\begin{table*}[ht]
\centering
\begin{tabular}{ r | r | r | r | r | l}
\toprule
\textbf{CWE} & \textbf{Train (\%)}  & \textbf{Test \#} & \textbf{TPR (\%)} & \textbf{FPR (\%)} &  \\
\midrule
CWE-119 & \textbf{15.16} & 313 & \textbf{39.30} & 3.55 & Improper Restriction of Operations within the Bounds of a Memory Buffer \\
CWE-120 & 2.29 & 49 & \textbf{40.82} & 3.55 & Buffer Copy without Checking Size of Input (`Classic Buffer Overflow') \\
CWE-125 & \textbf{11.08} & 239 & 27.20 & 3.55 & Out-of-bounds Read \\
CWE-189 & 2.97 & 57 & \textbf{31.58} & 3.55 & Numeric Errors \\
CWE-190 & \textbf{4.77} & 100 & 21.00 & 3.55 & Integer Overflow or Wraparound \\
CWE-200 & \textbf{5.10} & 131 & \textbf{31.30} & 3.55 & Exposure of Sensitive Information to an Unauthorized Actor \\
CWE-20 & \textbf{10.76} & 224 & \textbf{32.59} & 3.55 & Improper Input Validation \\
CWE-22 & 1.13 & 20 & 25.00 & 3.55 & Improper Limitation of a Pathname to a Restricted Directory (`Path Traversal') \\
CWE-264 & 3.55 & 73 & 28.77 & 3.55 & Permissions, Privileges, and Access Controls \\
CWE-269 & 1.14 & 23 & 8.70 & 3.55 & Improper Privilege Management \\
CWE-276 & 0.19 & 3 & 0 & 3.55 & Incorrect Default Permissions \\
CWE-284 & 3.35 & 77 & 25.97 & 3.55 & Improper Access Control \\
CWE-287 & 0.58 & 10 & 10.00 & 3.55 & Improper Authentication \\
CWE-306 & 0.00 & 0 & N/A & 3.55 & Missing Authentication for Critical Function \\
CWE-310 & 1.95 & 44 & 25.00 & 3.55 & Cryptographic Issues \\
CWE-352 & 0.10 & 1 & 0 & 3.55 & Cross-Site Request Forgery (CSRF) \\
CWE-362 & 2.62 & 61 & 16.39 & 3.55 & Race Condition \\
CWE-369 & 1.26 & 31 & 29.03 & 3.55 & Divide By Zero \\
CWE-399 & \textbf{5.29} & 110 & \textbf{41.82} & 3.55 & Resource Management Errors \\
CWE-400 & 2.38 & 34 & 5.88 & 3.55 & Uncontrolled Resource Consumption \\
CWE-401 & 1.83 & 33 & 24.24 & 3.55 & Missing Release of Memory after Effective Lifetime \\
CWE-415 & 1.55 & 30 & 30.00 & 3.55 & Double Free \\
CWE-416 & \textbf{5.46} & 112 & 17.86 & 3.55 & Use After Free \\
CWE-434 & 0.07 & 1 & 0 & 3.55 & Unrestricted Upload of File with Dangerous Type \\
CWE-476 & \textbf{5.00} & 106 & 17.92 & 3.55 & NULL Pointer Dereference \\
CWE-502 & 0.05 & 3 & \textbf{66.67} & 3.55 & Deserialization of Untrusted Data \\
CWE-611 & 0.09 & 3 & 0 & 3.55 & Improper Restriction of XML External Entity Reference \\
CWE-703 & \textbf{6.39} & 133 & 10.53 & 3.55 & Improper Check or Handling of Exceptional Conditions \\
CWE-77 & 0.18 & 6 & 16.67 & 3.55 & Command Injection \\
CWE-78 & 0.38 & 7 & 0 & 3.55 & OS Command Injection \\
CWE-787 & \textbf{15.57} & 311 & \textbf{33.76} & 3.55 & Out-of-bounds Write \\
CWE-79 & 0.47 & 12 & \textbf{50.00} & 3.55 & Cross-site Scripting \\
CWE-798 & 0.01 & 0 & N/A & 3.55 & Use of Hard-coded Credentials \\
CWE-862 & 0.26 & 6 & 16.67 & 3.55 & Missing Authorization \\
CWE-89 & 0.31 & 9 & \textbf{33.33} & 3.55 & SQL Injection \\
CWE-918 & 0.02 & 4 & 0 & 3.55 & Server-Side Request Forgery (SSRF) \\
CWE-94 & 0.69 & 15 & 0 & 3.55 & Improper Control of Generation of Code (`Code Injection') \\
\bottomrule
\end{tabular}
\caption{We evaluate the prediction performance of the CodeT5 Base model across top-25 CWEs and 12 most popular CWEs in \ours{}. We highlight the 10 highest training sample percentages and 10 highest TPR numbers in bold. Having more training samples for a specific CWE does not necessarily improve the model's prediction performance, and some CWEs are harder to learn than others. Most CWEs with 0\% TPR have under 10 samples in the test set.}
\label{tab:cwe-results}
\end{table*}

\begin{table*}[t!]
	\centering
	\begin{tabular}{c | r | r  r  r }
		\toprule
		\multirow{3}{*}{\textbf{Dataset}} & \multirow{3}{*}{\textbf{Correct Label}} & \multicolumn{3}{|c}{\textbf{Wrong Label}} \\
        & & \multirow{2}{*}{\begin{tabular}{@{}c@{}}{Vulnerability Spread}\\{Across Multiple Functions}\end{tabular}} & \multirow{2}{*}{Relevant Consistency} & \multirow{2}{*}{Irrelevant} \\
        & & & & \\
        \midrule
        \ours{} & 60\% & 10\% & 12\% & 18\% \\
        CVEFixes $\cup$ BigVul $\cup$ CrossVul & 36\% & 12\% & 12\% & 40\% \\
        CVEFixes & 51.7\% & 10.3\% & 17.3\% & 20.7\% \\
        BigVul & 25\% & 15.6\% & 9.4\% & 50\% \\
        CrossVul & 47.8\% & 13\% & 21.8\% & 17.4\% \\ 
        \bottomrule
	\end{tabular} 
	\caption{Label accuracy of four datasets, evaluated on a random sample of vulnerable functions.}
	\label{tab:labelerror}
\end{table*}

Figure~\ref{fig:class_weights} shows the gap between the F1 score on seen projects vs unseen projects for two CodeT5 Small models, one fine tuned with no weighting scheme and one fine tuned with class weights for cross entropy loss.
From the bars, we observe that using class weights reduces the gap between F1 score on seen vs unseen projects, with slight improvement to F1 score on seen projects and significant improvement for unseen projects. This means that using class weights improves the performance of the model over samples drawn from the same distribution as well as from a different distribution of new projects. However, there is still a large gap between 49.9\% F1 on seen projects vs 17.21\% F1 on unseen projects. As future research directions of the generalization problem, there is a lot of potential to further improve the model's performance over unknown projects.

\subsection{Performance on CWEs}

To understand the difficulty of learning different CWEs, we select 37 CWEs to examine the CodeT5 Base model's prediction performance when it is trained on Previous + \ours{}. The 37 CWEs include the top-25 CWEs according to MITRE~\cite{mitre}, and the 12 most common CWEs in \ours{} outside the top 25.
We select vulnerable functions belonging to these 37 CWEs and all nonvulnerable functions from the Previous + \ours{} test set obtained from the random split in Section~\ref{sec:add-our-data}.

\textbf{Result 10: Some CWEs are easier to learn than others regardless of the training data size.}
Table~\ref{tab:cwe-results} shows the CodeT5 Base model's prediction performance across the 37 CWEs. We have highlighted the 10 most prevalent CWEs in the training set and 10 highest True Positive Rate (TPR) numbers in bold. Note that all CWEs have the same False Positive Rate (FPR) since FPR is only related to nonvulnerable functions. We observe that having more samples for a particular CWE in the training set does not necessarily result in the model learning it better than CWEs with fewer training samples. Moreover, some CWEs with very few training samples are well-detected by the model. For example, CWE-502, CWE-79, CWE-89, all of which account for less than 2\% of the training data, have the highest TPRs. This suggests that some CWEs are easier to learn and do not require a large amount of training data, while others are more challenging to learn, even with more training samples. For instance, CWE-416 had 5.46\% of the training samples, but its TPR was only 17.86\%.

For some CWEs, we do not have enough test samples, resulting in extremely low TPR numbers. The ``Test \#'' column shows the number of vulnerable functions belonging to that CWE in the test set. For CWEs with 0\% TPR, most have less than 10 samples in the test set.

\section{Label Error Analysis}

While our dataset is designed to be as accurate as possible, some functions may be labelled erroneously.
To label vulnerable functions, we follow the methodology used in Devign~\cite{zhou2019devign}, \reveal{}~\cite{chakraborty2021deep}, BigVul~\cite{fan2020ac}, CrossVul~\cite{nikitopoulos2021crossvul}, and CVEFixes~\cite{bhandari2021cvefixes}, which considers a function vulnerable if it was changed by a commit that is identified as fixing a vulnerability, based on security issue trackers.
Although our labeling technique is state-of-the-art and can scale effectively, we cannot guarantee that every function changed by each such commit is vulnerable, so some labels may be inaccurate.

To quantify the amount of label noise as a result of this labeling methodology, we manually assess the accuracy of labels for the \ours{}, CVEFixes, BigVul, and CrossVul datasets. Among previous datasets, we chose CVEFixes, BigVul, and CrossVul because they provide the commit ID that changed the vulnerable function, which allows us to verify whether a function is vulnerable in that specific version of the project.

We randomly sample 50 vulnerable functions from \ours{}, and 50 vulnerable functions from the union of previous three datasets (CVEFixes $\cup$ BigVul $\cup$ CrossVul). Then, we manually analyze whether the vulnerable function has the correct label or wrong label.
We inform this decision by examining the code of the function labelled vulnerable, both before and after the commit, the commit it was supposedly fixed in, the CVE description, and developer discussions in the security issue tracker.
We confirm a function as correctly labelled vulnerable if the vulnerability exists in that function, and is not spread across multiple functions.
We observed three categories of label errors: 1) the vulnerability is spread across multiple functions, 2) the function is not vulnerable, but changing the function is relevant to fixing the vulnerability (e.g., to adjust calling parameters), and 3) the function is not vulnerable and irrelevant to the vulnerability (e.g., a vulnerability-fixing commit changes the spaces in some nonvulnerable functions, or makes irrelevant functionality changes to nonvulnerable functions).

Table~\ref{tab:labelerror} shows our analysis results. The vulnerable function labels are 60\% accurate in \ours{}, which is 24 percentage points higher than the previous three datasets (CVEFixes $\cup$ BigVul $\cup$ CrossVul). Within these three datasets, CVEFixes is the most accurate one, whereas BigVul has very low label accuracy, only 25\%. We observe that many commits included in BigVul from the Chromium and Android projects are not relevant to fixing vulnerabilities at all. We also found that the percentage of irrelevant functions is surprisingly high, ranging from 17.4\% to 50\% in four datasets. These functions are not related to the vulnerability, but since they were changed by the vulnerability-fixing commits, the automatic labeling process labels them as vulnerable.


Concurrent work also examined label noise and also found significant label errors in the BigVul and Devign datasets~\cite{croft2023data}.
Compared to their categorization,
we have a stricter criteria to label a function as vulnerable: we consider the caller of a vulnerable function as non-vulnerable; they considered it vulnerable.
Also, if a function is only part of the vulnerability, and if the vulnerability cannot be recognized from the code of this function alone, we consider that a wrong label; they considered it correct.
Taking into account the differences in categorization, our findings for BigVul (the only dataset common to their and our work) are largely consistent with their findings.

\section{Limitations}


The label noise in our dataset and prior datasets may introduce errors into our measurement of the performance of all models on the test set.
We hope that releasing our dataset will enable the community to explore methods to remediate the effects of label noise in the future.

In retrospect, the de-duplication procedure in our dataset and prior datasets could be improved.
As part of the label noise analysis, we discovered that 4\% of \ours{} labels and 6\% of (CVEFixes $\cup$ BigVul $\cup$ CrossVul) labels were erroneous because the commit made whitespace-only changes to some functions, and these were treated as security fixes during labelling.
Therefore, normalizing the whitespace in all functions before de-duplication could slightly improve label accuracy, and might have other benefits.

There is a risk of contamination, i.e., test data leaking into pre-training data, as LLMs are pre-trained on text and code, which could conceivably include blog articles or code patches related to security vulnerabilities included in our test set.
Many of our models (CodeBERT, GraphCodeBERT, PolyCoder, CodeT5 Small, CodeT5 Base, NatGen) were only pre-trained on code, not on other text or code changes, so
could have been exposed to code in our test set but
were unlikely to be exposed to a description of which code is vulnerable.
This could potentially affect our results in ways that we cannot measure.
Other models (RoBERTa, GPT-2 Base, CodeGPT, T5 Base) were pre-trained on text, and so could possibly have been exposed to blog articles that describe vulnerable source code.
We suspect that this is very rare, but we cannot measure it, so we cannot rule out the possibility of test set contamination.
The latter models (RoBERTa, GPT-2 Base, CodeGPT, T5 Base) performed relatively poorly in our experiment in any case.

There is also a risk that cloned code could cause test set contamination, if the cloned code was subsequently modified slightly (thus evading our de-duplication efforts).

\section{Conclusion}

This paper presents a new dataset, \ours{}, for detecting software vulnerabilities using deep learning. The dataset contains 18,945 vulnerable functions spanning 155 CWEs and 330,492 nonvulnerable functions, extracted from 7,514 commits, which is more diverse and twice the size of the previous largest and most diverse dataset, CVEFixes. We use this new dataset to study the effectiveness of various deep learning architectures in detecting vulnerabilities. We have experimented with 11 different deep learning architectures from four model families: Graph Neural Networks (GNN), RoBERTa, GPT-2, and T5.
The results suggest that the increased diversity and volume of training data examined in this paper is beneficial for vulnerability detection, especially for large language models, but it is unclear whether even larger datasets would help or not.
Code-specific pretraining tasks appear to be a promising research direction for deep learning based vulnerability detection.
Our results highlight a major challenge for future research: improving deep learning models so they generalize to unknown projects.
We release the \ours{} dataset to the community at \url{https://github.com/wagner-group/diversevul}.


\begin{acks}
We are grateful to Bryce Casaje for his contributions exploring multiple approaches for dataset construction, including manual labelling of commits, automated text-based labelling, and more. We are grateful to Kexin Pei for his advice on large language model fine tuning.
We gratefully acknowledge the anonymous reviewers for many helpful remarks that significantly improved the paper.
This research was supported by the NSF under grant CNS-2154873,
by the joint KACST - UC Berkeley Center of Excellence for Secure Computing,
by C3.AI's Digital Transformation Institute, and by the Center for AI
Safety Compute Cluster.  Any opinions, findings, and conclusions
or recommendations expressed in this material are those of the
author(s) and do not necessarily reflect the views of the sponsors.
\end{acks}

%
%

\bibliographystyle{ACM-Reference-Format}
\bibliography{vulref}


\begin{thebibliography}{33}


\ifx \showCODEN    \undefined \def \showCODEN     #1{\unskip}     \fi
\ifx \showDOI      \undefined \def \showDOI       #1{#1}\fi
\ifx \showISBNx    \undefined \def \showISBNx     #1{\unskip}     \fi
\ifx \showISBNxiii \undefined \def \showISBNxiii  #1{\unskip}     \fi
\ifx \showISSN     \undefined \def \showISSN      #1{\unskip}     \fi
\ifx \showLCCN     \undefined \def \showLCCN      #1{\unskip}     \fi
\ifx \shownote     \undefined \def \shownote      #1{#1}          \fi
\ifx \showarticletitle \undefined \def \showarticletitle #1{#1}   \fi
\ifx \showURL      \undefined \def \showURL       {\relax}        \fi
\providecommand\bibfield[2]{#2}
\providecommand\bibinfo[2]{#2}
\providecommand\natexlab[1]{#1}
\providecommand\showeprint[2][]{arXiv:#2}

\bibitem[Bessey et~al\mbox{.}(2010)]%
        {bessey10}
\bibfield{author}{\bibinfo{person}{Al Bessey}, \bibinfo{person}{Ken Block},
  \bibinfo{person}{Ben Chelf}, \bibinfo{person}{Andy Chou},
  \bibinfo{person}{Seth~Hallem Bryan~Fulton}, \bibinfo{person}{Charles
  Henri-Gros}, \bibinfo{person}{Asya Kamsky}, \bibinfo{person}{Scott McPeak},
  {and} \bibinfo{person}{Dawson Engler}.} \bibinfo{year}{2010}\natexlab{}.
\newblock \showarticletitle{A few billion lines of code later: using static
  analysis to find bugs in the real world}.
\newblock \bibinfo{journal}{\emph{Commun. ACM}} \bibinfo{volume}{53},
  \bibinfo{number}{2} (\bibinfo{date}{February} \bibinfo{year}{2010}).
\newblock


\bibitem[Bhandari et~al\mbox{.}(2021)]%
        {bhandari2021cvefixes}
\bibfield{author}{\bibinfo{person}{Guru Bhandari}, \bibinfo{person}{Amara
  Naseer}, {and} \bibinfo{person}{Leon Moonen}.}
  \bibinfo{year}{2021}\natexlab{}.
\newblock \showarticletitle{CVEfixes: automated collection of vulnerabilities
  and their fixes from open-source software}. In
  \bibinfo{booktitle}{\emph{Proceedings of the 17th International Conference on
  Predictive Models and Data Analytics in Software Engineering}}.
  \bibinfo{pages}{30--39}.
\newblock


\bibitem[Chakraborty et~al\mbox{.}(2022)]%
        {chakraborty2022natgen}
\bibfield{author}{\bibinfo{person}{Saikat Chakraborty},
  \bibinfo{person}{Toufique Ahmed}, \bibinfo{person}{Yangruibo Ding},
  \bibinfo{person}{Premkumar~T Devanbu}, {and} \bibinfo{person}{Baishakhi
  Ray}.} \bibinfo{year}{2022}\natexlab{}.
\newblock \showarticletitle{NatGen: generative pre-training by
  “naturalizing” source code}. In \bibinfo{booktitle}{\emph{Proceedings of
  the 30th ACM Joint European Software Engineering Conference and Symposium on
  the Foundations of Software Engineering}}. \bibinfo{pages}{18--30}.
\newblock


\bibitem[Chakraborty et~al\mbox{.}(2021)]%
        {chakraborty2021deep}
\bibfield{author}{\bibinfo{person}{Saikat Chakraborty}, \bibinfo{person}{Rahul
  Krishna}, \bibinfo{person}{Yangruibo Ding}, {and} \bibinfo{person}{Baishakhi
  Ray}.} \bibinfo{year}{2021}\natexlab{}.
\newblock \showarticletitle{Deep learning based vulnerability detection: Are we
  there yet}.
\newblock \bibinfo{journal}{\emph{IEEE Transactions on Software Engineering}}
  (\bibinfo{year}{2021}).
\newblock


\bibitem[Challande et~al\mbox{.}(2022)]%
        {challande2022building}
\bibfield{author}{\bibinfo{person}{Alexis Challande}, \bibinfo{person}{Robin
  David}, {and} \bibinfo{person}{Gu{\'e}na{\"e}l Renault}.}
  \bibinfo{year}{2022}\natexlab{}.
\newblock \showarticletitle{Building a Commit-level Dataset of Real-world
  Vulnerabilities}. In \bibinfo{booktitle}{\emph{Proceedings of the Twelveth
  ACM Conference on Data and Application Security and Privacy}}.
  \bibinfo{pages}{101--106}.
\newblock


\bibitem[Corporation(2023)]%
        {mitre}
\bibfield{author}{\bibinfo{person}{The~MITRE Corporation}.} \bibinfo{year}{Last
  accessed on March 28, 2023}\natexlab{}.
\newblock \bibinfo{title}{2022 CWE Top 25 Most Dangerous Software Weaknesses}.
\newblock
\newblock
\urldef\tempurl%
\url{https://cwe.mitre.org/top25/archive/2022/2022_cwe_top25.html}
\showURL{%
\tempurl}


\bibitem[Croft et~al\mbox{.}(2023)]%
        {croft2023data}
\bibfield{author}{\bibinfo{person}{Roland Croft}, \bibinfo{person}{M~Ali
  Babar}, {and} \bibinfo{person}{Mehdi Kholoosi}.}
  \bibinfo{year}{2023}\natexlab{}.
\newblock \showarticletitle{Data quality for software vulnerability datasets}.
  In \bibinfo{booktitle}{\emph{2023 IEEE/ACM 45th International Conference on
  Software Engineering (ICSE)}}.
\newblock


\bibitem[Devlin et~al\mbox{.}(2018)]%
        {devlin2018bert}
\bibfield{author}{\bibinfo{person}{Jacob Devlin}, \bibinfo{person}{Ming-Wei
  Chang}, \bibinfo{person}{Kenton Lee}, {and} \bibinfo{person}{Kristina
  Toutanova}.} \bibinfo{year}{2018}\natexlab{}.
\newblock \showarticletitle{Bert: Pre-training of deep bidirectional
  transformers for language understanding}.
\newblock \bibinfo{journal}{\emph{arXiv preprint arXiv:1810.04805}}
  (\bibinfo{year}{2018}).
\newblock


\bibitem[Fan et~al\mbox{.}(2020)]%
        {fan2020ac}
\bibfield{author}{\bibinfo{person}{Jiahao Fan}, \bibinfo{person}{Yi Li},
  \bibinfo{person}{Shaohua Wang}, {and} \bibinfo{person}{Tien~N Nguyen}.}
  \bibinfo{year}{2020}\natexlab{}.
\newblock \showarticletitle{AC/C++ code vulnerability dataset with code changes
  and CVE summaries}. In \bibinfo{booktitle}{\emph{Proceedings of the 17th
  International Conference on Mining Software Repositories}}.
  \bibinfo{pages}{508--512}.
\newblock


\bibitem[Feng et~al\mbox{.}(2020)]%
        {feng2020codebert}
\bibfield{author}{\bibinfo{person}{Zhangyin Feng}, \bibinfo{person}{Daya Guo},
  \bibinfo{person}{Duyu Tang}, \bibinfo{person}{Nan Duan},
  \bibinfo{person}{Xiaocheng Feng}, \bibinfo{person}{Ming Gong},
  \bibinfo{person}{Linjun Shou}, \bibinfo{person}{Bing Qin},
  \bibinfo{person}{Ting Liu}, \bibinfo{person}{Daxin Jiang}, {et~al\mbox{.}}}
  \bibinfo{year}{2020}\natexlab{}.
\newblock \showarticletitle{Codebert: A pre-trained model for programming and
  natural languages}.
\newblock \bibinfo{journal}{\emph{arXiv preprint arXiv:2002.08155}}
  (\bibinfo{year}{2020}).
\newblock


\bibitem[Guo et~al\mbox{.}(2020)]%
        {guo2020graphcodebert}
\bibfield{author}{\bibinfo{person}{Daya Guo}, \bibinfo{person}{Shuo Ren},
  \bibinfo{person}{Shuai Lu}, \bibinfo{person}{Zhangyin Feng},
  \bibinfo{person}{Duyu Tang}, \bibinfo{person}{Shujie Liu},
  \bibinfo{person}{Long Zhou}, \bibinfo{person}{Nan Duan},
  \bibinfo{person}{Alexey Svyatkovskiy}, \bibinfo{person}{Shengyu Fu},
  {et~al\mbox{.}}} \bibinfo{year}{2020}\natexlab{}.
\newblock \showarticletitle{Graphcodebert: Pre-training code representations
  with data flow}.
\newblock \bibinfo{journal}{\emph{arXiv preprint arXiv:2009.08366}}
  (\bibinfo{year}{2020}).
\newblock


\bibitem[Husain et~al\mbox{.}(2019)]%
        {husain2019codesearchnet}
\bibfield{author}{\bibinfo{person}{Hamel Husain}, \bibinfo{person}{Ho-Hsiang
  Wu}, \bibinfo{person}{Tiferet Gazit}, \bibinfo{person}{Miltiadis Allamanis},
  {and} \bibinfo{person}{Marc Brockschmidt}.} \bibinfo{year}{2019}\natexlab{}.
\newblock \showarticletitle{Codesearchnet challenge: Evaluating the state of
  semantic code search}.
\newblock \bibinfo{journal}{\emph{arXiv preprint arXiv:1909.09436}}
  (\bibinfo{year}{2019}).
\newblock


\bibitem[Li et~al\mbox{.}(2015)]%
        {li2015gated}
\bibfield{author}{\bibinfo{person}{Yujia Li}, \bibinfo{person}{Daniel Tarlow},
  \bibinfo{person}{Marc Brockschmidt}, {and} \bibinfo{person}{Richard Zemel}.}
  \bibinfo{year}{2015}\natexlab{}.
\newblock \showarticletitle{Gated graph sequence neural networks}.
\newblock \bibinfo{journal}{\emph{arXiv preprint arXiv:1511.05493}}
  (\bibinfo{year}{2015}).
\newblock


\bibitem[Li et~al\mbox{.}(2021)]%
        {li2021sysevr}
\bibfield{author}{\bibinfo{person}{Zhen Li}, \bibinfo{person}{Deqing Zou},
  \bibinfo{person}{Shouhuai Xu}, \bibinfo{person}{Hai Jin},
  \bibinfo{person}{Yawei Zhu}, {and} \bibinfo{person}{Zhaoxuan Chen}.}
  \bibinfo{year}{2021}\natexlab{}.
\newblock \showarticletitle{Sysevr: A framework for using deep learning to
  detect software vulnerabilities}.
\newblock \bibinfo{journal}{\emph{IEEE Transactions on Dependable and Secure
  Computing}} \bibinfo{volume}{19}, \bibinfo{number}{4} (\bibinfo{year}{2021}),
  \bibinfo{pages}{2244--2258}.
\newblock


\bibitem[Li et~al\mbox{.}(2018)]%
        {li2018vuldeepecker}
\bibfield{author}{\bibinfo{person}{Zhen Li}, \bibinfo{person}{Deqing Zou},
  \bibinfo{person}{Shouhuai Xu}, \bibinfo{person}{Xinyu Ou},
  \bibinfo{person}{Hai Jin}, \bibinfo{person}{Sujuan Wang},
  \bibinfo{person}{Zhijun Deng}, {and} \bibinfo{person}{Yuyi Zhong}.}
  \bibinfo{year}{2018}\natexlab{}.
\newblock \showarticletitle{Vuldeepecker: A deep learning-based system for
  vulnerability detection}.
\newblock \bibinfo{journal}{\emph{arXiv preprint arXiv:1801.01681}}
  (\bibinfo{year}{2018}).
\newblock


\bibitem[Liu et~al\mbox{.}(2019)]%
        {liu2019roberta}
\bibfield{author}{\bibinfo{person}{Yinhan Liu}, \bibinfo{person}{Myle Ott},
  \bibinfo{person}{Naman Goyal}, \bibinfo{person}{Jingfei Du},
  \bibinfo{person}{Mandar Joshi}, \bibinfo{person}{Danqi Chen},
  \bibinfo{person}{Omer Levy}, \bibinfo{person}{Mike Lewis},
  \bibinfo{person}{Luke Zettlemoyer}, {and} \bibinfo{person}{Veselin
  Stoyanov}.} \bibinfo{year}{2019}\natexlab{}.
\newblock \showarticletitle{Roberta: A robustly optimized bert pretraining
  approach}.
\newblock \bibinfo{journal}{\emph{arXiv preprint arXiv:1907.11692}}
  (\bibinfo{year}{2019}).
\newblock


\bibitem[Lu et~al\mbox{.}(2021)]%
        {lu2021codexglue}
\bibfield{author}{\bibinfo{person}{Shuai Lu}, \bibinfo{person}{Daya Guo},
  \bibinfo{person}{Shuo Ren}, \bibinfo{person}{Junjie Huang},
  \bibinfo{person}{Alexey Svyatkovskiy}, \bibinfo{person}{Ambrosio Blanco},
  \bibinfo{person}{Colin Clement}, \bibinfo{person}{Dawn Drain},
  \bibinfo{person}{Daxin Jiang}, \bibinfo{person}{Duyu Tang}, {et~al\mbox{.}}}
  \bibinfo{year}{2021}\natexlab{}.
\newblock \showarticletitle{Codexglue: A machine learning benchmark dataset for
  code understanding and generation}.
\newblock \bibinfo{journal}{\emph{arXiv preprint arXiv:2102.04664}}
  (\bibinfo{year}{2021}).
\newblock


\bibitem[Mirsky et~al\mbox{.}(2023)]%
        {mirskyvulchecker}
\bibfield{author}{\bibinfo{person}{Yisroel Mirsky}, \bibinfo{person}{George
  Macon}, \bibinfo{person}{Michael Brown}, \bibinfo{person}{Carter Yagemann},
  \bibinfo{person}{Matthew Pruett}, \bibinfo{person}{Evan Downing},
  \bibinfo{person}{Sukarno Mertoguno}, {and} \bibinfo{person}{Wenke Lee}.}
  \bibinfo{year}{2023}\natexlab{}.
\newblock \showarticletitle{VulChecker: Graph-based Vulnerability Localization
  in Source Code}. In \bibinfo{booktitle}{\emph{USENIX Security 2023}}.
\newblock


\bibitem[Nikitopoulos et~al\mbox{.}(2021)]%
        {nikitopoulos2021crossvul}
\bibfield{author}{\bibinfo{person}{Georgios Nikitopoulos},
  \bibinfo{person}{Konstantina Dritsa}, \bibinfo{person}{Panos Louridas}, {and}
  \bibinfo{person}{Dimitris Mitropoulos}.} \bibinfo{year}{2021}\natexlab{}.
\newblock \showarticletitle{CrossVul: a cross-language vulnerability dataset
  with commit data}. In \bibinfo{booktitle}{\emph{Proceedings of the 29th ACM
  Joint Meeting on European Software Engineering Conference and Symposium on
  the Foundations of Software Engineering}}. \bibinfo{pages}{1565--1569}.
\newblock


\bibitem[of~Standards and Technology(2023a)]%
        {nvd}
\bibfield{author}{\bibinfo{person}{National~Institute of Standards} {and}
  \bibinfo{person}{Technology}.} \bibinfo{year}{Last accessed on March 19,
  2023}\natexlab{a}.
\newblock \bibinfo{title}{National Vulnerability Database}.
\newblock
\newblock
\urldef\tempurl%
\url{https://nvd.nist.gov/}
\showURL{%
\tempurl}


\bibitem[of~Standards and Technology(2023b)]%
        {sard}
\bibfield{author}{\bibinfo{person}{National~Institute of Standards} {and}
  \bibinfo{person}{Technology}.} \bibinfo{year}{Last accessed on March 19,
  2023}\natexlab{b}.
\newblock \bibinfo{title}{NIST Software Assurance Reference Dataset}.
\newblock
\newblock
\urldef\tempurl%
\url{https://samate.nist.gov/SARD}
\showURL{%
\tempurl}


\bibitem[Okun et~al\mbox{.}(2013)]%
        {okun2013report}
\bibfield{author}{\bibinfo{person}{Vadim Okun}, \bibinfo{person}{Aurelien
  Delaitre}, \bibinfo{person}{Paul~E Black}, {et~al\mbox{.}}}
  \bibinfo{year}{2013}\natexlab{}.
\newblock \showarticletitle{Report on the static analysis tool exposition
  (sate) iv}.
\newblock \bibinfo{journal}{\emph{NIST Special Publication}}
  \bibinfo{volume}{500} (\bibinfo{year}{2013}), \bibinfo{pages}{297}.
\newblock


\bibitem[Radford et~al\mbox{.}(2019)]%
        {radford2019language}
\bibfield{author}{\bibinfo{person}{Alec Radford}, \bibinfo{person}{Jeffrey Wu},
  \bibinfo{person}{Rewon Child}, \bibinfo{person}{David Luan},
  \bibinfo{person}{Dario Amodei}, \bibinfo{person}{Ilya Sutskever},
  {et~al\mbox{.}}} \bibinfo{year}{2019}\natexlab{}.
\newblock \showarticletitle{Language models are unsupervised multitask
  learners}.
\newblock \bibinfo{journal}{\emph{OpenAI blog}} \bibinfo{volume}{1},
  \bibinfo{number}{8} (\bibinfo{year}{2019}), \bibinfo{pages}{9}.
\newblock


\bibitem[Raffel et~al\mbox{.}(2020)]%
        {raffel2020exploring}
\bibfield{author}{\bibinfo{person}{Colin Raffel}, \bibinfo{person}{Noam
  Shazeer}, \bibinfo{person}{Adam Roberts}, \bibinfo{person}{Katherine Lee},
  \bibinfo{person}{Sharan Narang}, \bibinfo{person}{Michael Matena},
  \bibinfo{person}{Yanqi Zhou}, \bibinfo{person}{Wei Li}, {and}
  \bibinfo{person}{Peter~J Liu}.} \bibinfo{year}{2020}\natexlab{}.
\newblock \showarticletitle{Exploring the limits of transfer learning with a
  unified text-to-text transformer}.
\newblock \bibinfo{journal}{\emph{The Journal of Machine Learning Research}}
  \bibinfo{volume}{21}, \bibinfo{number}{1} (\bibinfo{year}{2020}),
  \bibinfo{pages}{5485--5551}.
\newblock


\bibitem[Russell et~al\mbox{.}(2018)]%
        {russell2018automated}
\bibfield{author}{\bibinfo{person}{Rebecca Russell}, \bibinfo{person}{Louis
  Kim}, \bibinfo{person}{Lei Hamilton}, \bibinfo{person}{Tomo Lazovich},
  \bibinfo{person}{Jacob Harer}, \bibinfo{person}{Onur Ozdemir},
  \bibinfo{person}{Paul Ellingwood}, {and} \bibinfo{person}{Marc McConley}.}
  \bibinfo{year}{2018}\natexlab{}.
\newblock \showarticletitle{Automated vulnerability detection in source code
  using deep representation learning}. In \bibinfo{booktitle}{\emph{2018 17th
  IEEE international conference on machine learning and applications (ICMLA)}}.
  IEEE, \bibinfo{pages}{757--762}.
\newblock


\bibitem[Steenhoek et~al\mbox{.}(2023)]%
        {steenhoek2022empirical}
\bibfield{author}{\bibinfo{person}{Benjamin Steenhoek},
  \bibinfo{person}{Md~Mahbubur Rahman}, \bibinfo{person}{Richard Jiles}, {and}
  \bibinfo{person}{Wei Le}.} \bibinfo{year}{2023}\natexlab{}.
\newblock \showarticletitle{An empirical study of deep learning models for
  vulnerability detection}. In \bibinfo{booktitle}{\emph{2023 IEEE/ACM 45th
  International Conference on Software Engineering (ICSE)}}.
\newblock


\bibitem[Thapa et~al\mbox{.}(2022)]%
        {thapa2022transformer}
\bibfield{author}{\bibinfo{person}{Chandra Thapa}, \bibinfo{person}{Seung~Ick
  Jang}, \bibinfo{person}{Muhammad~Ejaz Ahmed}, \bibinfo{person}{Seyit
  Camtepe}, \bibinfo{person}{Josef Pieprzyk}, {and} \bibinfo{person}{Surya
  Nepal}.} \bibinfo{year}{2022}\natexlab{}.
\newblock \showarticletitle{Transformer-Based Language Models for Software
  Vulnerability Detection}. In \bibinfo{booktitle}{\emph{Proceedings of the
  38th Annual Computer Security Applications Conference}}.
  \bibinfo{pages}{481--496}.
\newblock


\bibitem[Wang et~al\mbox{.}(2021a)]%
        {wang2021patchdb}
\bibfield{author}{\bibinfo{person}{Xinda Wang}, \bibinfo{person}{Shu Wang},
  \bibinfo{person}{Pengbin Feng}, \bibinfo{person}{Kun Sun}, {and}
  \bibinfo{person}{Sushil Jajodia}.} \bibinfo{year}{2021}\natexlab{a}.
\newblock \showarticletitle{Patchdb: A large-scale security patch dataset}. In
  \bibinfo{booktitle}{\emph{2021 51st Annual IEEE/IFIP International Conference
  on Dependable Systems and Networks (DSN)}}. IEEE, \bibinfo{pages}{149--160}.
\newblock


\bibitem[Wang et~al\mbox{.}(2021b)]%
        {wang2021codet5}
\bibfield{author}{\bibinfo{person}{Yue Wang}, \bibinfo{person}{Weishi Wang},
  \bibinfo{person}{Shafiq Joty}, {and} \bibinfo{person}{Steven~CH Hoi}.}
  \bibinfo{year}{2021}\natexlab{b}.
\newblock \showarticletitle{Codet5: Identifier-aware unified pre-trained
  encoder-decoder models for code understanding and generation}.
\newblock \bibinfo{journal}{\emph{arXiv preprint arXiv:2109.00859}}
  (\bibinfo{year}{2021}).
\newblock


\bibitem[Xu et~al\mbox{.}(2022)]%
        {xu2022systematic}
\bibfield{author}{\bibinfo{person}{Frank~F Xu}, \bibinfo{person}{Uri Alon},
  \bibinfo{person}{Graham Neubig}, {and} \bibinfo{person}{Vincent~J
  Hellendoorn}.} \bibinfo{year}{2022}\natexlab{}.
\newblock \showarticletitle{A Systematic Evaluation of Large Language Models of
  Code}.
\newblock \bibinfo{journal}{\emph{arXiv preprint arXiv:2202.13169}}
  (\bibinfo{year}{2022}).
\newblock


\bibitem[Yamaguchi et~al\mbox{.}(2014)]%
        {yamaguchi2014modeling}
\bibfield{author}{\bibinfo{person}{Fabian Yamaguchi}, \bibinfo{person}{Nico
  Golde}, \bibinfo{person}{Daniel Arp}, {and} \bibinfo{person}{Konrad Rieck}.}
  \bibinfo{year}{2014}\natexlab{}.
\newblock \showarticletitle{Modeling and discovering vulnerabilities with code
  property graphs}. In \bibinfo{booktitle}{\emph{2014 IEEE Symposium on
  Security and Privacy}}. IEEE, \bibinfo{pages}{590--604}.
\newblock


\bibitem[Zheng et~al\mbox{.}(2021)]%
        {zheng2021d2a}
\bibfield{author}{\bibinfo{person}{Yunhui Zheng}, \bibinfo{person}{Saurabh
  Pujar}, \bibinfo{person}{Burn Lewis}, \bibinfo{person}{Luca Buratti},
  \bibinfo{person}{Edward Epstein}, \bibinfo{person}{Bo Yang},
  \bibinfo{person}{Jim Laredo}, \bibinfo{person}{Alessandro Morari}, {and}
  \bibinfo{person}{Zhong Su}.} \bibinfo{year}{2021}\natexlab{}.
\newblock \showarticletitle{D2A: a dataset built for AI-based vulnerability
  detection methods using differential analysis}. In
  \bibinfo{booktitle}{\emph{2021 IEEE/ACM 43rd International Conference on
  Software Engineering: Software Engineering in Practice (ICSE-SEIP)}}. IEEE,
  \bibinfo{pages}{111--120}.
\newblock


\bibitem[Zhou et~al\mbox{.}(2019)]%
        {zhou2019devign}
\bibfield{author}{\bibinfo{person}{Yaqin Zhou}, \bibinfo{person}{Shangqing
  Liu}, \bibinfo{person}{Jingkai Siow}, \bibinfo{person}{Xiaoning Du}, {and}
  \bibinfo{person}{Yang Liu}.} \bibinfo{year}{2019}\natexlab{}.
\newblock \showarticletitle{Devign: Effective vulnerability identification by
  learning comprehensive program semantics via graph neural networks}.
\newblock \bibinfo{journal}{\emph{Advances in neural information processing
  systems}}  \bibinfo{volume}{32} (\bibinfo{year}{2019}).
\newblock


\end{thebibliography}

\appendix
\section{Model Training Setups}
\label{appendix:training_details}

\subsection{\reveal{} Setup}
We use Joern on GitHub~\footnote{After commit a6aa08ee9842eedb52e149695e3a34500b6ceab0 on Oct 11, 2022.} to obtain the Code Property Graphs. This is a newer version than what \reveal{} used, because if we use the same old version of Joern as in the \reveal{} paper, almost half of the functions in all datasets cannot be extracted into graphs.

For the Gated Graph Neural Network, we set maximum training epochs to be 50 for Previous + \ours{} dataset and 100 for Previous dataset, and pick the model with the best validation F1 score, for experiments in Section~\ref{sec:add-our-data}. We set maximum training epochs to be 60 for experiments in Section~\ref{sec:generalization}. We follow the original setting in \reveal{} source code to use Adam optimizer with learning rate 0.0001, and weight decay 0.001.

To train the classification layers in \reveal{}, we set the maximum number of epochs to be 100 and follow authors’ set up: we stop the training procedure if F1-score on validation set does not increase in 5 epochs. We follow the original setting in \reveal{} source code to use Adam optimizer with learning rate 0.001, and no weight decay.

\subsection{Fine Tuning Setup}

To fine tune LLM models, we apply a linear classification head over the Tranformer model, following standard methods. For RoBERTa, CodeBERT, and GraphCodeBERT, we apply the linear layer over the embedding that represents the first token ([CLS]). For GPT-2-Base, CodeGPT, and PolyCoder, we apply the linear layer over the embedding of the last token. For the T5 Base, CodeT5 Small, CodeT5 Base, and NatGen, we apply the linear layer over the embeddings of the last decoder state.

We use training batch size 32, learning rate 2e-5, Adam optimizer, and train for 10 epochs. We use a linear learning rate decay with warm up of 1,000 steps. We check the model's validation performance every 1,000 steps, and save the model with the best validation performance for testing. We use the same learning rate for all models and all training data setups with one exception. When we train RoBERTa on Previous + \ours{} from the random data split (in Section~\ref{sec:add-our-data}), we use learning rate 1e-5, since a larger learning rate results in a degenerate model that always predicts a function as nonvulnerable.




\end{document}